\documentclass[11pt]{article}	
\usepackage{amsmath,amsthm,amsfonts,amscd, amsmath} 
\usepackage{amssymb,amsfonts,amsmath,amscd,amsthm}	
\usepackage{graphicx}
\usepackage[square,numbers]{natbib}
\usepackage{setspace}
\usepackage[caption=false]{subfig}  
\usepackage{verbatim}   	
\usepackage{makeidx}    	
\usepackage{color}
\usepackage{enumerate}
\usepackage{ upgreek }
\usepackage{authblk}
\usepackage{epsfig}     	
\usepackage{url}		
\usepackage{tcolorbox}
\usepackage{multirow}

\usepackage[pdftex,bookmarks]{hyperref}

\graphicspath{{../figs/}} 


\theoremstyle{definition}

\theoremstyle{remark}

\usepackage[toc,page]{appendix}

\newcommand{\latexe}{{\LaTeX\kern.125em2%
           \lower.5ex\hbox{$\varepsilon$}}}

\chardef\bslash=`\\	

\makeatletter		
\def\square{\RIfM@\bgroup\else$\bgroup\aftergroup$\fi
 \vcenter{\hrule\hbox{\vrule\@height.6em\kern.6em\vrule}%
                       \hrule}\egroup}
                                               
\makeatother		
\makeindex  

\usepackage{fancyhdr}
\begin{document}
\title{
A Predictive Discrete‐Continuum Multiscale Model of Plasticity With Quantified Uncertainty
}

\author[1]{Jingye Tan}
\author[2]{Umberto Villa}
\author[3]{Nima Shamsaei}
\author[3]{Shuai Shao}
\author[4]{Hussein M. Zbib}
\author[1]{Danial Faghihi\thanks{corresponding author, danialfa@buffalo.edu (D. Faghihi)}}
\affil[1]{Department of Mechanical and Aerospace Engineering, University at Buffalo}
\affil[2]{Electrical and Systems Engineering, Washington University in St. Louis}
\affil[3]{Department of Mechanical Engineering, Auburn University}
\affil[4]{School of Mechanical and Materials Engineering, Washington State University}
\renewcommand\Authands{ and }

\maketitle


\vspace{-0.3in}
\begin{abstract}

Multiscale models of materials, consisting of upscaling discrete simulations to continuum models, are unique in their capability to simulate complex materials behavior. The fundamental limitation in multiscale models is the presence of uncertainty in the computational predictions delivered by them. 
In this work, a sequential multiscale model has been developed, incorporating discrete dislocation dynamics (DDD) simulations and a strain gradient plasticity (SGP) model to predict the size effect in plastic deformations of metallic micro-pillars.
The DDD simulations include uniaxial compression of micro-pillars with different sizes and over a wide range of initial dislocation densities and spatial distributions of dislocations. 
{An} SGP model is employed at the continuum level that accounts for the size-dependency of flow stress and hardening rate. 
Sequences of uncertainty analyses have been performed to assess the predictive capability of the multiscale model. 
{The variance-based global sensitivity analysis determines the effect of parameter uncertainty on the SGP model prediction.}
The multiscale model is then constructed by calibrating the continuum model using the data furnished by the DDD simulations. A Bayesian calibration method is implemented to quantify the uncertainty due to microstructural randomness in discrete dislocation simulations (density and spatial distributions of dislocations) on the macroscopic continuum model prediction (size effect in plastic deformation).  
The outcomes of this study indicate that the discrete-continuum multiscale model can accurately simulate the plastic deformation of micro-pillars, despite the significant uncertainty in the DDD results.
Additionally, depending on the macroscopic features represented by the DDD simulations, the SGP model can reliably predict the size effect in plasticity responses of the micropillars with below 10\% of error.

\end{abstract}


\section{Introduction}

{
Small volume metallic components, with the size in the range of the material intrinsic length scale, exhibit distinctive plastic deformation responses compared to macroscale materials. 
Computational models that can characterize the properties of these materials, such as thin-films and micro-pillars, and enable predictions of the fabrication processes are critically needed to advance the miniaturization of microsystem technology. 
One example of such fabrication processes is micro-scale metal forming, e.g., \cite{Kuhfuss2020, meng2013, zhang2020understanding}, which relies on plastic deformation for the manufacturing of metallic microstructures with desirable geometries.
In addition to micro-fabrication's technology, investigating the mechanical behavior of small-volume materials provides a unique opportunity to understand the underlying micromechanics of deformation and failure processes in larger-scale materials and structures \cite{voyiadjis2013localization, voyiadjis2010}. 
Experimental studies of plasticity in small volume metals using instrumented nanoindentation \cite{chen2001, wang2004}, microtorsion \cite{fleck1994strain}, and tensile testing \cite{espinosa2004, xiang2005, xiang2006} indicate that the dislocation interactions and surface/interface effects play substantial roles in micro-scale plasticity responses. 
A common micro-scale mechanical test is micro-pillar compression that provides an approximately uniform stress state. Meng et al. \cite{meng2014micro, meng2014, meng2017} proposed a protocol for testing substrate interfacial failures through compression loading of micro-pillars in the axial direction, allowing failure stress measurements to be acquired from the same metal under different geometries.}
Such microstructural interactions result in significant dependency of plastic behavior to the size of the material, in which smaller structure shows higher strength, fatigue resistance, and fracture resistance \cite{faghihi2012indent, faghihi2010size, XIAO2019216, LE201956, RAN201365, WU201836, LYU201746}. The size effect in plastic deformation of the microscale materials is attributed to the geometrically necessary dislocations (GNDs) caused by the crystal lattice's geometrical constraints. The evolution of GNDs gives rise to the deformation resistance by acting as obstacles to the statistically stored dislocations (SSDs) \cite{arsenlis1999}. Another source of the size effect in microscale plasticity is the strengthening due to the scarcity of dislocations in small volumes, and the impacts of surfaces and obstacles such as grain boundaries \cite{zbib1998plastic, akarapu2010, nicola2006, ElAwady2011}.


Numerous computational models address the size effect in plasticity responses of microscale metallic materials.
High-fidelity discrete models, such as molecular dynamics 
\cite{voyiadjis2016, yaghoobi2016, li2020, KRASNIKOV2020102672, XIONG201286}
and discrete dislocation dynamics (DDD) \cite{GiessenNeedleman1995, ElAwady2009, ElAwady2015nature, MCELFRESH2020102848, NIU2019262, PRASTITI2020102615, JIANG2019111}, explicitly simulate the microstructural evolutions leading to plastic deformation.
On the other hand, crystal plasticity \cite{gurtin2010, gurtinphno2011, bargmann2011, zikry1996, shi2009, SEDIGHIANI2020102779, CASTELLUCCIO2017} 
and strain gradient plasticity models \cite{alrubvoyiadjis2006, hutchinson1997, gudmundson2019, JEBAHI2020, SUN2019215, BITTENCOURT2019280, BITTENCOURT2018169, LUBARDA201626, BITTENCOURT202020} relies on a continuum description of the overall microstructural behavior resulting in the size effect phenomena.
The multiscale method bridging discrete simulations and continuum models alleviates the computational burden of the discrete model and enables simulations of large materials systems on practical time and length scales
\cite{BERTIN2019268, ELTERS2019257, KRASNIKOV201921, XU201591, XIONG201533}. 
{An attractive class of computational plasticity models is DDD} \cite{GiessenNeedleman1995, shu2001, Rao2010, ElAwady2015}, which simulates the motion of individual dislocations using the elastic theory of dislocations and enables explicit modeling of the dislocation interactions with other defects, particles, and internal/external surfaces. 
Since the microscopic material behavior is captured naturally within the dynamics of the microstructural evolutions \cite{grohzbib2009}, DDD directly accounts for the effect of the material intrinsic length scale on the plastic responses.
As a result, DDD simulations have emerged as a powerful computational tool in investigating microscale plasticity phenomena, including lattice and grain boundary effects \cite{yang2015, ElAwady2015}, dislocation network \cite{shao2018frequency}, the behavior of thin films \cite{vsivska2009, bittencourt2003, wang2014, shao2020}, and dislocation boundaries \cite{khan2014}. 
Despite these advances, the high computational costs of discrete dislocation simulations have limited the applicability of these models to simple geometries and length scales of 10$\mu m$ and time scales of 1$ms$ \cite{sills2016}. 
Another class of models that address microscale plasticity, {without limitation in length and time scales}, is the strain gradient plasticity (SGP). After the initial introduction of the gradient theory of plasticity by Aifantis \cite{aifantis1987, aifantis1984}, there have been extensive developments of SGP models by Fleck and Hutchinson \cite{FleckHutchinson1993}, and Nix and Gao \cite{nixgao1998}, along with the theories developed within a thermodynamically consistent framework found in \cite{fleck2009i, fleck2009ii, gudmundson2019, gudmundson2004, gurtinReddy2014, gurtinanand2005ii, gurtinanand2005i, voyiadjisrub2007, anand2005, nicola2005, Steinmann2019}. 
For a comprehensive review of the theoretical developments of SGP models and their applications, the interested readers are referred to Voyiadjis and Song \cite{voyiadjis2019}, and the references cited therein.
Such progressions in the SGP models, along with their numerical analyses \cite{ reddy2013, garikipati2003}, {finite element method solutions} \cite{DeBorst1996, mcbride2009}, and analytical and experimental interpretations of microstructural length scale \cite{faghihi2010indent, faghihi2010size, faghihi2012indent, DAHLBERG2019220, LIU2017156, ABUALRUB20041139} reveal the strength of SGP in depicting the plastic deformation in micro-scale materials. 
{SGP models have been widely employed to model thin films tension and shear \cite{voyiadjis2012ijp, FLECK20012245, gudmundson2004} along with simulating the micropillar compression \cite{zhang2014, meng2014micro, HUSSER2014206}.
Despite the ability to simulate complex geometry and lower computational costs of the SGP models, determining the model parameters from observational data is challenging and prevent the broad application of SGP models in predicting complex material responses \cite{meng2014micro}.}


Potentially predictive computational tools to address micro-scale plasticity behavior of materials are multiscale models, beginning with DDD and hierarchical transitioning to continuum SGP models. The promise of such a multiscale model is that the continuum model, informed by high-fidelity discrete simulations, can efficiently characterize and simulate the materials and structural responses. 
There are several attempts in the literature to integrate the DDD simulations with the continuum plasticity models \cite{bayerschen2015, lyuzbib2016, vsivska2009, shu2001, bittencourt2003, GiessenNeedleman1997}. These modeling approaches have provided further insight into understanding the underlying microstructural behavior responsible for plasticity responses of micro-scale materials. 
The main challenge in multiscale modeling of materials is the uncertainty in computational prediction delivered by them. Predictive computational modeling is emerging in current literature to describe the use of data science tools to integrate observational data into physics-based computational models and enhance the predictive power of the models along with quantifying uncertainty \cite{odenbabuska2017, oden2010computer}. The sources of uncertainty in discrete-continuum multiscale modeling include the microstructural randomness, the inherent stochasticity of discrete models, the inadequacy of continuum model in depicting microstructural phenomena, and the loss of information through the discrete to continuum upscaling \cite{oden2018adaptive, kouchmeshky2010}. An example of uncertainty in multiscale models of plastic deformation is the stochastic nature of the onset of plasticity due to the variation of the initial dislocation content and heterogeneity, as observed in microscale experiments \cite{ngan2010, ng2008, zaiser2008}. Such stochasticity in flow stress and hardening responses can be simulated by DDD methods, while the SGP model, based on deterministic plasticity theories, fail to address such randomness. The presence of uncertainties in multiscale modeling gives rise to the need for uncertainty quantification (UQ) method to evaluate the sensitivity of responses of interest to variation of material microstructure at different scales and propagation of uncertainty through multiscale model chains \cite{TALLMAN2020113009, WHELAN2019106673, WANG20201, McDowell2020, TALLMAN2020449, panchal2013, najm2011}.


In this contribution, we develop a discrete to continuum multiscale framework for predicting the plastic responses of micro-pillars under compression with quantified uncertainty. The high-fidelity discrete model is based on the multiscale dislocation dynamics plasticity (MDDP) simulations. The MDDP simulations are conducted on the metallic micro-pillars of {heights} 200 nm to 1000 nm under compression and over a wide range of initial dislocation densities and spatial distributions of dislocations. 
In the continuum level, we use an SGP model, including energetic and dissipative length scales, that account for size effect in both flow stress and hardening rate.
{A notable feature of this study is the comprehensive treatment of uncertainty in the multiscale models arising from the randomness in initial dislocation density and distributions, as well as the SGP modeling errors.} Variance-based global sensitivity analyses are conducted to describe the impact of parameters' variabilities and the micro-pillar size on the SGP model prediction. We then use a Bayesian model calibration framework to determine the SGP model's parameters using the synthetic data furnished by MDDP simulations. In particular, we assess the effect of the microstructural randomness (density and spatial distributions of dislocations) on the macroscopic SGP model prediction (size effect in plastic deformation). Various training and testing data sets are taken into account to explore how well the SGP parameters are learned from MDDP simulations and investigate the SGP model's ability in predicting the micro-pillar responses.


The rest of this manuscript is organized as follows. Section 2 presents a summary of the discrete and continuum models used in the multiscale framework.
The uncertainty quantification methods consist of global sensitivity analyses, Bayesian inference, and forward model prediction under uncertainty are described in Section 3. The results of uncertainty analyses, calibration, and the multiscale model's prediction are provided in section 4. The Discussion and Conclusions are given in Sections 5 and 6.

\section{Computational Models of Microscale Plasticity}

The sequential multiscale model of plastic deformation, in this work, consists of upscaling a discrete dislocation dynamics model to a continuum strain-gradient plasticity model. 
Both models can capture the unique plastic behavior in small-volume materials.
The discrete (high fidelity) model provides detailed microstructural evolutions with high computational costs. 
In contrast, the continuum (low fidelity) model captures the macroscopic responses with no limitation in length and time scales. 
In this section, we summarize the basics of each computational model.

\subsection{Discrete dislocation dynamics model.}\label{sec:mddp}

The discrete dislocation simulations of this work rely on the multiscale dislocation dynamics plasticity (MDDP) developed by Zbib and his co-workers \cite{zbib2002, hiratanizbib2002, zbib2003}. 
This hybrid multiscale model couples a 3D stochastic dislocation dynamic with continuum plasticity such that the discrete simulations replace the macroscopic constitutive equations.  
An overview of MDDP framework is provided in this section and more details can be found in \cite{zbib2002, hiratanizbib2002, grohzbib2009, zbib2011, zbibshehadeh2005, zbib2003}.

The MDDP model simulates the dynamics of the microstructure based on the elastic interactions among dislocations.
{The curved dislocation lines are represented 
by the discrete segments moving on a lattice and}
the dynamics of each dislocation segment is obtained by the Newton equation of motion \cite{grohzbib2009}, 
\begin{equation}\label{eq:motion}
m_s \dot{{v}}_i + B {{v}}_i = {F}^*,
\end{equation}
where ${v}_i$ is the velocity of the $i$-th dislocation, $m_s$ is the effective mass per unit length of dislocation line, and $B$ is the drag coefficient (inverse of dislocation mobility). The glide force vector ${F}^*$ per unit length arises from various components, including dislocation-dislocation interaction, self force (line tension) external load, dislocation-obstacle interaction, and thermal forces.
The effective resolved shear stress on the slip systems, $\tau^*$, is
\begin{eqnarray}
\tau^*  & = & | \tau^{**} | - \tau_{\text{friction}},
\end{eqnarray}
where 
$\tau^{**}$ represents the driving force and 
$\tau_{\text{friction}}$ is
the resolved shear stress corresponding to the 
friction force.
Neglecting the inertia term, the equation of motion (\ref{eq:motion}) can then be rewritten as,
\begin{equation}\label{eq:motion_shear}
{v}_i=\left\{\begin{matrix}
{0} & \text{if} & \tau^*\leq 0\\ 
\text{sign}(\tau^{**})\dfrac{\tau^*{b}_i}{B} & \text{if} & \tau^*> 0,
\end{matrix}\right.
\end{equation}
where $b_i$ is the Burgers vector's magnitude \cite{zbib2002, zbib2003}.
Once the velocity of each dislocation is computed from (\ref{eq:motion_shear}), MDDP uses a numerical algorithm to check the possible interactions between dislocations { such as annihilation, Hirth lock, Glissile junction, and Lomer Lock \cite{grohzbib2009}.}
In MDDP, the increment of plastic shear $\delta \gamma_i^p$, due to the gliding of dislocation $i$, is governed by Orowan’s law,
\begin{equation}
\delta \gamma_i^p = \frac{b_i  \, \delta\! A_i}{V}
\end{equation}
where $\delta A_i$ is the area swept during gliding, and $V$ is the volume of the sheared body. 
Having the increment of plastic shear on slip system $k$, the {increment of} plastic strain rate tensor components, for Face Centered Cubic (FCC) systems, are computed using,
\begin{equation}\label{eq:dd:plasticinc}
\delta \dot{\boldsymbol{\epsilon}}^p = \sum_{k=1}^{12} \frac{1}{2} \left( \mathbf{n}^{(k)} \otimes \mathbf{l}^{(k)} + \mathbf{n}^{(k)} \otimes \mathbf{l}^{(k)} \right) \delta {\dot{\gamma}^{p(k)}},
\end{equation}
where $\mathbf{n}^{(k)}$ and $\mathbf{l}^{(k)}$ are the unit vectors parallel
to the slip plane normal and the Burgers vector, respectively.
{
To integrate the dislocation dynamics with a continuum plasticity, 
MDDP uses the plastic strain-rate tensor in (\ref{eq:dd:plasticinc}) within the incremental Hooke’s law used to evaluate the Cauchy stress $\mathbf{T}$,
\begin{equation}
\delta \dot{\mathbf{T}} = \mathcal{C} (\delta\dot{\boldsymbol{\epsilon}} - \delta\dot{\boldsymbol{\epsilon}}^p),
\end{equation}
where $ \mathcal{C}$ being the fourth-order elastic tensor, and $\delta\dot{\boldsymbol{\epsilon}}$ is the rate of incremental strain in the continuum level. 
}

{
MDDP is successfully employed in investigating microscale plasticity phenomena, including free surface effects, micro-shear bands, dislocation boundaries, and deformation of thin layers; see, e.g., \cite{zbib2011} and the reference therein. 
Despite the advantages of MDDP in simulating small volume plasticity phenomena, the computational cost of this model is scaled with dislocation density. Simulating complex micro-manufacturing processes or deformation of the macro-scale metallic components is computationally infeasible using MDDP and requires continuum models that are informed and validated using the MDDP simulations.
}

\subsection{Strain gradient plasticity model}\label{sec:sgp}
The continuum model of plasticity in microscale material, in this work, relies on a strain-gradient plasticity theory, developed by Faghihi and Voyiadjis in a series of works, e.g., \cite{faghihi2012phd, voyiadjis2014, faghihi2012thermal, faghihi2014jemt}. Built on Gurtin and co-workers' earlier work, e.g., \cite{gurtinanand2005i, gurtinReddy2014, gurtinphno2011}, Faghihi and Voyiadjis developed a thermodynamically consistent framework for fully coupled temperature and rate-dependent strain gradient plasticity, including internal interfaces and a generalized heat conduction model.
Later, Song and Voyiadjis extended the thermo-mechanical gradient theory by the finite deformations formulation and addressed the non-proportional stretch-passivation loading problem \cite{ song2020, song2018, voyiadjis2017book, voyiadjis2017}.
Consistently with the MDDP simulations of the micro-pillar compression,
here we discard the thermal and interface effects of the general thermo-mechanical SGP model.
A summary of the reduced SGP model is laid down in this section, and detailed information about the complete model can be found in \cite{voyiadjis2012ijp, voyiadjis2013, voyiadjis2013pm,faghihi2013, voyiadjis2014ijss}.

Let the reference configuration $\Omega$ be a bounded 
Lipschitz domain in $\mathbb{R}^3$, 
$\mathbf{x}$ denoting the position of a material point, and
$\Gamma_D \cup \Gamma_N =\partial \Omega$ and $\Gamma_H \cup \Gamma_F =\partial \Omega$ be non-overlapping boundaries.
The displacement gradient $\nabla\mathbf{u}$ is decomposed into elastic and plastic parts according to,
\begin{equation}
\nabla\mathbf{u} = \mathbf{H}^e + \mathbf{H}^p,
\end{equation}
where the elastic counterpart of the displacement gradient, $\mathbf{H}^e$,
represents the recoverable rotation and stretching of the material,
while $\mathbf{H}^p$ represents the plastic distortion
and describes the evolution of dislocations and other defects in the material. 
Under the small deformation and plasticity irrotational assumptions, 
the infinitesimal strain, $\boldsymbol{\epsilon}$, 
the elastic strain, $\boldsymbol{\epsilon}^e$, and 
plastic strain, $\boldsymbol{\epsilon}^p$, are given by,
\begin{equation}
\boldsymbol{\epsilon} = \frac{1}{2} \left(\nabla\mathbf{u} + (\nabla\mathbf{u})^T\right),\quad
\boldsymbol{\epsilon}^p = \boldsymbol{\epsilon} - \boldsymbol{\epsilon}^e = \frac{1}{2} \left(\mathbf{H}^p + (\mathbf{H}^p)^T\right), \quad
\text{tr} \boldsymbol{\epsilon}^p = 0.
\end{equation}
To account for strain-gradient effects due to the presence of geometrically necessary dislocations (GNDs), we make use of the Nye’s dislocation density tensor \cite{nye1953},
\begin{equation}
\mathbf{G} = \text{Curl} \; \boldsymbol{\epsilon}^p, 
\end{equation}
where the magnitude of this tensor is related to total GNDs density through the relation $\| \mathbf{G} \| = b\rho_G$, with $b$ being the magnitude of the Burgers vector and $\rho_G$  being the GNDs density. 

Balance equations of the continuum theory are derived from the principle of virtual power written in terms of kinematic quantities:
the elastic strain rate $\dot{\boldsymbol{\epsilon}^e}$,
plastic strain rate $\dot{\boldsymbol{\epsilon}^p}$, and 
gradient of plastic strain rate $\nabla \dot{\boldsymbol{\epsilon}^p}$. 
The principle of virtual power results in
the \textit{macro-force balance},
\begin{equation}
\nabla \cdot \mathbf{T} + \mathbf{f} = \mathbf{0}, \quad \text{in}\; \Omega,
\end{equation}
together with \textit{macroscopic boundary conditions},
\begin{equation}\label{eq:macroforce}
 \mathbf{T} {\mathbf{n}_N} = \mathbf{t} \quad \text{on}\; \Gamma_N, 
 \quad\quad
 \mathbf{u} = \mathbf{u}^\dagger \quad \text{on}\; \Gamma_D,  
\end{equation}
where $\mathbf{T}$ is the Cauchy stress tensor, 
$\mathbf{f}$ is the body force, 
$\mathbf{t}$ is the surface traction, 
$\mathbf{u}^\dagger$ is a prescribed displacement on the boundary $\Gamma_D$,
 and {$\mathbf{n}_N$} in the outward unit normal on the boundary $\Gamma_N$.
Additionally, the \textit{microforce balance} is given by {\cite{voyiadjis2014ijss, faghihi2014jemt}}
\begin{equation}\label{eq:microforce}
\mathbf{T}_0 - \mathbf{R} + \nabla\cdot\mathbf{S} = \mathbf{0}, \quad \text{in}\; \Omega,
\end{equation}
together with \textit{microscopic boundary conditions}
\begin{equation}
 \mathbf{S} {\mathbf{n}_F} = \mathbf{m} \quad \text{on}\; \Gamma_F,
 \quad\quad
 \boldsymbol{\epsilon}^p = {\boldsymbol{\epsilon}^p}^\dagger \quad \text{on}\; \Gamma_H,  
\end{equation}
where $\mathbf{T}_0$ is the deviatoric part of the Cauchy stress,
the second-order tensor $\mathbf{R}$ is the microforces conjugate to plastic strain, 
and the third-order tensor $\boldsymbol{S}$ is the microforces conjugate to the plastic strain gradient.
The prescribed plastic strain on the boundary $\Gamma_H$ is denoted by ${\boldsymbol{\epsilon}^p}^\dagger$, {$\mathbf{n}_F$ in the outward unit normal on the boundary $\Gamma_F$}, and 
$\mathbf{m}$ is called the micro-traction giving rise to interface (grain boundary) models \cite{faghihi2012thermal}.
Following the thermodynamical derivations, the Clausius-Duhem inequality of the SGP model is expressed as {\cite{voyiadjis2014ijss, faghihi2014jemt}},
\begin{equation}\label{eq:CD}
 \mathbf{T} : \dot{\boldsymbol{\epsilon}^e} + \mathbf{R} : \dot{\boldsymbol{\epsilon}^p} + \mathbf{S} : \nabla \dot{\boldsymbol{\epsilon}^p} - \dot{\Psi} \geq 0,
\end{equation}
where the free energy $\Psi$ takes the form,
\begin{equation}
\Psi = \Psi^e + \Psi^d.
\end{equation}
The elastic component of the free energy is,
\begin{equation}
\Psi^e = \frac{1}{2} (\boldsymbol{\epsilon} - \boldsymbol{\epsilon}^p): \mathcal{C} (\boldsymbol{\epsilon} - \boldsymbol{\epsilon}^p),
\end{equation}
where $\mathcal{C} = 2\mu\boldsymbol{\epsilon} + \lambda [\text{tr}\boldsymbol{\epsilon}]\mathbf{I}$ is the fourth-order elastic tensor, with $\mu$ and $\lambda$ being the (positive) Lam\'{e} constants.
The well-known relation of the elastic modulus $E$ with the Lam\'{e} constant is $E = \mu (3\lambda+2\mu)/(\lambda+\mu)$.
The defect energy $\Psi^d$ accounts for the interaction among slip systems (characterized by the accumulated plastic strain $p = \| \boldsymbol{\epsilon}^p \| $ in macroscale)
as well as the short-range interactions between GNDs moving on close slip planes (represented by plastic strain gradient). 
Accordingly, the constitutive equation of the defect free energy is postulated as,
\begin{equation}\label{eq:defect_energy}
\Psi^d = \frac{h}{r} \left( \exp(-rp) + rp \right) + 
\frac{1}{2} \mu \ell_{\text{en}}^2 \| \nabla \boldsymbol{\epsilon}^p\|^2.
\end{equation}
The first term of the $\Psi^d$ represents the forest dislocations leading to isotropic hardening, in which $h$ and $r$ are the hardening parameters. The second term in (\ref{eq:defect_energy}) accounts for the core energy of GNDs leading to a backstress and hence to kinematic hardening \cite{gurtinanand2005i, gurtinanand2009FHA}, where $ \| \nabla \boldsymbol{\epsilon}^p\|^2 = \nabla \boldsymbol{\epsilon}^p: \nabla \boldsymbol{\epsilon}^p$. 
The parameter $ \ell_{\text{en}}$ is the energetic length scale that controls the 
non-local short range interaction among GNDs.
The Cauchy stress and the energetic counterparts of the microstresses are given by,
\begin{eqnarray} \label{eq:energetics}
\mathbf{T} & = & \frac{\partial \Psi}{\partial \boldsymbol{\epsilon}^e} = \mathcal{C} (\boldsymbol{\epsilon} - \boldsymbol{\epsilon}^p),\nonumber\\
\mathbf{R}_{\text{en}} & = & \frac{\partial \Psi}{\partial \boldsymbol{\epsilon}^p} = h- h\exp(-rp), \nonumber\\
\mathbf{S}_{\text{en}} & = & \frac{\partial \Psi}{\partial \nabla \boldsymbol{\epsilon}^p} = \mu \ell_{\text{en}}^2 \nabla \boldsymbol{\epsilon}^p.
\end{eqnarray}
Next, substituting (\ref{eq:energetics}) into (\ref{eq:CD}), leads to the reduced dissipation inequality {\cite{voyiadjis2013pm}},
\begin{equation}
\Phi = \mathbf{R}_{\text{dis}} : \dot{\boldsymbol{\epsilon}}^p + \mathbf{S}_{\text{dis}} : \nabla \dot{\boldsymbol{\epsilon}}^p \geq 0.
\end{equation}
This inequality is the foundation for constructing a plasticity flow rule, where $\Phi$ is the energy dissipation potential.
The constitutive equations for the dissipative thermodynamical stresses are {(see \cite{voyiadjis2014} for details)},
\begin{eqnarray} \label{eq:dissipative}
\mathbf{R}_{\text{dis}} & = & \frac{\partial \Phi}{\partial \dot{\boldsymbol{\epsilon}}^p} = Y \left( \frac{\dot{\wp}}{q} \right)^m \frac{\dot{\boldsymbol{\epsilon}}^p}{\dot{\wp}} , \nonumber\\
\mathbf{S}_{\text{dis}} & = & \frac{\partial \Phi}{\partial \nabla \dot{\boldsymbol{\epsilon}}^p} = Y \ell^2_{\text{dis}} \left( \frac{\dot{\wp}}{q} \right)^m \frac{\nabla \dot{\boldsymbol{\epsilon}}^p}{\dot{\wp}},
\end{eqnarray}
where $Y$ is a macroscopic measure of the initial slip resistance known as yield strength,
$\ell_{\text{dis}}$ is the dissipative length scale controlling the gradient strengthening, i.e., size dependent flow stress.
The strain rate dependencies of the SGP model are governed by the power law, where $m$ and $q$ are visco-plastic parameters.
Also, the effective nonlocal flow rate $\dot{\wp}$ is defined as,
\begin{equation}
\dot{\wp} = \sqrt{\|\dot{\boldsymbol{\epsilon}}^p \|^2 + \ell_{\text{dis}}^2 \| \nabla \dot{\boldsymbol{\epsilon}}^p \|^2 },
\end{equation}
that is a phenomenological relation describing the energy dissipation due to the motion of both SSDs and GNDs
\cite{gurtinanand2005i, gudmundson2019}.

The micro-force balance (\ref{eq:microforce}), equipped by the constitutive relations for $\mathbf{R} = \mathbf{R}_{\text{en}} + \mathbf{R}_{\text{dis}}$ and $\mathbf{S} = \mathbf{S}_{\text{en}} + \mathbf{S}_{\text{dis}}$, can be written as a second-order partial-differential
equation for the plastic strain,
\begin{equation}
\mathbf{T}_0 - \left[ -\mu \ell^2_{\text{en}} \Delta \boldsymbol{\epsilon}^p\right] = 
h(1- \exp(-rp))
 + Y \left( \frac{\dot{\wp}}{q} \right)^m \frac{\dot{\boldsymbol{\epsilon}}^p}{\dot{\wp}}
 - Y \ell^2_{\text{dis}} \nabla \cdot \left[ \left( \frac{\dot{\wp}}{q} \right)^m \frac{\nabla \dot{\boldsymbol{\epsilon}}^p}{\dot{\wp}}\right],
\end{equation}
where $\Delta = \nabla\cdot\nabla$ is the Laplacian operator and
we have written the term $\nabla \cdot \mathbf{S}_{\text{en}}$ on the left as it represents a backstress.

\subsubsection{Finite element solution of the strain gradient plasticity model}\label{sec:fem}

We simulate the compression of the micro-pillars with one-dimensional (1D) SGP model with the domain size of $L$ {(see Figure \ref{fig:pillar})}.
The governing equations of the model consist of macro-force and micro-force balances, and the corresponding boundary conditions become,
\begin{equation}
\left.\begin{matrix}
\nabla {T}  =  {0} \\ 
{T}_0 - {R} + \nabla{S}  =  {0}\\
u({y=}0) = 0, \quad u({y=}L) = u^\dagger\\
\epsilon^p({y=}0) =0, \quad  \epsilon^p({y=}L) = 0,
\end{matrix}\right\},
\end{equation}
where the body force is neglected from the macro-force balance and scalar variables are used in the 1D formulation.
For modeling the micro-pillar compression, quasi-static conditions are assumed, and the increment of compressive deformation at a constant strain rate is applied to one boundary ($u({y=}L) = u^\dagger$).
{The justifications of the employed boundary conditions are provided in section \ref{sec:forward}.}
We make use of a dual-mixed variational formulation of the coupled nonlinear system of equations, involving both displacement field ${u}$ and the plastic strain field ${\epsilon}^p$ as separate unknowns.
To define the relevant finite element space, we define a Hilbert space $\mathcal{Z} = \{{z} \in H^1(\Omega) : {z}|_{\Gamma} ={0} \}$.
The variational problem, considering {${z}$} and ${w}$ as test functions, is defined as: 
Find $({u}, {\epsilon}^p) \in \mathcal{Z} \times \mathcal{Z}$, for all ${z},{w} \in \mathcal{Z}$
\begin{equation}\label{eq:variational}
\left.\begin{matrix}
{\displaystyle \int_\Omega {T} \cdot \nabla {z} \; dy} & = & {0} 
\\ 
{\displaystyle \int_\Omega \left( ({R}-{T_0} ) {w} + {S} \cdot \nabla {w} \right) \; dy} & = & {0}
\end{matrix}\right\},
\end{equation}
The numerical solution of the dual-mixed variational form (\ref{eq:variational}) is obtained through continuous finite element methods.
To keep the compatibility of the dual-mixed finite element function space, second-order Lagrange is used for the displacement field $u$, while the plastic strain $\epsilon^p$ is discretized by the first-order Lagrange elements.
{
The numerical algorithm in this work is based on a Newton-type method for solving the system of nonlinear equations arises in (\ref{eq:variational}).
The Newton method's convergence rate is quadratic; however, its global convergence demands good initial guesses.
To this end, for the SGP model, the convergence is achieved by using numerical condition techniques on the prescribed boundary displacement $u^\dagger$. 
}

\section{Methods for Uncertainty Analyses}

Characterizing uncertainty in predictive computational models comprises two processes: 
(i) the statistical forward process, which involves propagating input uncertainties through the computational model to quantify the uncertainties in model output or the quantity of interests (QoI's);
(ii) the statistical inverse problem in which the probability densities of the models' uncertain parameters are estimated using observations.
This section presents the techniques and computational methods we use for uncertainty treatment in the MDDP to SGP sequential multi-scale model, including global sensitivity analysis, Bayesian statistical inference,
and forward uncertainty propagation.

\subsection{Variance-based global sensitivity analyses}\label{sec:sensitivity}

Global sensitivity analyses enable studying the impact of the randomness in a model's input parameters on the uncertainty of the model's outputs \cite{Sobol1990, SaltelliSobol1995, Saltelli2010, saltelli2008, saltelli2009}. 
We employ a variance-based global sensitivity analysis method \cite{Sobol1990, Sobol1993, Sobol2007},
in which the parameter sensitivity is calculated by the conditional variance in the model output caused by an input. 
A summary of this method is provided in this section.

Let $Q$ be a univariate output of a model (or a QoI) with $K$ uncertain parameters, $\boldsymbol{\theta} = \{\theta_1, \theta_2, \cdots, \theta_K\}$. 
One can write the following decomposition for the variance of the model output by conditioning with respect to all the input parameters but $\theta_k, \; k=1,2,\cdots, K$ \cite{HommaSaltelli1996, SaltelliTarantola2002},
\begin{equation}\label{eq:var_decomp}
\mathbb{V} (Q) = \mathbb{V}_{\boldsymbol{\theta}_{\sim k}} \left(\mathbb{E}_{\theta_k} (Q | \boldsymbol{\theta}_{\sim k}) \right) + \mathbb{E}_{\boldsymbol{\theta}_{\sim k}} \left(\mathbb{V}_{\theta_k} (Q | \boldsymbol{\theta}_{\sim k}) \right),
\end{equation}
where $\theta_k$ is the $k$-th input factor, 
$\boldsymbol{\theta}_{\sim k}$ shows the matrix of all parameters except $\theta_k$, and 
$\mathbb{E}_{\boldsymbol{\theta}_{\sim k}}(\cdot)$ and $\mathbb{V}_{\boldsymbol{\theta}_{\sim k}}(\cdot)$ indicate the mean and variance taken over all possible values of $\boldsymbol{\theta}_{\sim k}$ while $\theta_k$ is fixed.
{A variance based sensitivity measure is the \textit{total effect index} $S_k$ defined as (see \cite{HommaSaltelli1996, SaltelliTarantola2002} for more details),}
\begin{equation}\label{eq:ST}
S_{k} 
 = \frac{\mathbb{E}_{\boldsymbol{\theta}_{\sim k}} \left(\mathbb{V}_{\theta_k} (Q | \boldsymbol{\theta}_{\sim k}) \right)}{\mathbb{V}(Q)}
 = 1- \frac{\mathbb{V}_{\boldsymbol{\theta}_{\sim k}} \left(\mathbb{E}_{\theta_k} (Q | \boldsymbol{\theta}_{\sim k}) \right)}{\mathbb{V}(Q)}.
\end{equation}
In Eq. (\ref{eq:ST}), $\mathbb{V}_{\boldsymbol{\theta}_{\sim k}} \left(\mathbb{E}_{\theta_k} ( Q | \boldsymbol{\theta}_{\sim k}) \right)$ represents the expected variance reduction if all values other than $\theta_{k}$ are fixed and
$\mathbb{E}_{\boldsymbol{\theta}_{\sim k}} \left(\mathbb{V}_{\theta_k} (Q | \boldsymbol{\theta}_{\sim k}) \right)$ indicates the remaining variance of $Q$ for fixed $\theta_k$. 
Accordingly, the total effect index $S_{k}$ measures the impact of the input $\theta_k$ to the variation of the model output. A small total effect index for a parameter $\theta_k$ indicates that fixing that parameter at any value, within its uncertainty range, will not appreciably impact the model output. 
%

\subsubsection{Numerical estimator of total effect sensitivity index.}
To compute the total sensitivity index, we use an efficient Monte-Carlo estimator proposed by Saltelli \cite{Saltelli2002, saltelli2008, Saltelli2010, HommaSaltelli1996}. Estimating $S_k$ using this method consists of constructing two $N\times K$ matrices, 
$\mathbf{A}$ and $\mathbf{B}$, in which $N$ random samples are drawn from the probability distributions of the uncertain parameters.
The matrices $\mathbf{A}_{\mathbf{B}}^{(k)}, \; k=1,2, \cdots, K$ are constructed from all columns of $\mathbf{A}$ except the $k$th column, that comes from $\mathbf{B}$.
The model outputs for the input parameters in the each row of the matrices $\mathbf{A}$ and $\mathbf{A}_{\mathbf{B}}^{(k)}$ are stored in the vectors $\mathbf{y}_{\boldsymbol{A}}$ and $\mathbf{y}_{\boldsymbol{A}\boldsymbol{B}}^{(k)}$, respectively.
The total-effect index for parameter $ {\theta}_k$, is then approximated using the following estimator \cite{Saltelli2010},
\begin{equation}\label{eq:STapprox}
S_{k} \approx \frac{1}{2N}\sum_{j=1}^{N}\left(\left(\mathbf{y}_{\boldsymbol{A}}\right)_j-\left(\mathbf{y}_{\boldsymbol{A}\boldsymbol{B}}^{(k)}\right)_j\right)^2.
\end{equation}
The Monte-Carlo estimator of $S_k$ in (\ref{eq:STapprox}) decreases the cost of estimating multi-dimensional integrals from $N^2$ to $N(K+2)$ model evaluation \cite{Saltelli2010}.

\subsection{Bayesian inference for model calibration}\label{sec:bayes}

An essential process in predictive modeling of complex physical processes is to calibrate the model's parameters using a set of observational data and check the validity of the model. The assessment of prediction reliability consists of characterizing the uncertainties in the model parameters and data and propagating such uncertainty to the QoI as the computational prediction target. 
The sources of uncertainty are the error in the computational model in depicting the physical reality and noise and variabilities in data \cite{odenbabuska2017}.
In recent years, Bayesian approaches to statistical inference problems have been gaining popularity in broad areas of material science and engineering \cite{singer2013, honarmandi2020}. Such momentum is because these methods offer general frameworks for predictive modeling while providing means to portray uncertainty. 
Here, we summarize our Bayesian calibration process, as described in \cite{odenbabuska2017} and implemented in \cite{prudencio2015, prudencio2014CompB, farrell2015jcp, oden2015amses, faghihi2018fatigue} for predictive modeling of various physical systems.

Consider $\boldsymbol{\theta}$ to be a vector of model parameters and 
$\mathbf{D}$ to be the observational (training) data.
In the Bayesian setting, 
$\boldsymbol{\theta}$ and $\mathbf{D}$ are random variables represented by probability density functions (PDFs), $\pi(\boldsymbol{\theta})$ and $\pi(\mathbf{D})$. 
The calibration process enables one to identify model parameters that can explain the data $\mathbf{D}$. 
To characterize the uncertainties in both the data and the model parameters, we make use of a statistical inference method in which probability
density functions of the calibrated parameters are given by Bayes' theorem \cite{jaynes2003}:
\begin{equation}\label{eq:Bayes_calib}
\pi_{\text{post}}({\boldsymbol{\theta}}|\mathbf{D})
=
\frac
{\pi_{\text{like}}(\mathbf{D}|{\boldsymbol{\theta}})\cdot \pi_{\text{prior}}({\boldsymbol{\theta}})}
{\pi_{\text{evid}}(\mathbf{D})}.
\end{equation}
In Eq. \eqref{eq:Bayes_calib},
$\pi_{\text{post}}({\boldsymbol{\theta}}|\mathbf{D})$ is the posterior PDF defining the Bayesian update of the prior information represented by $\pi_{\text{prior}}(\boldsymbol{\theta})$,
$\pi_{\text{like}}(\mathbf{D}|\boldsymbol{\theta})$ is the likelihood PDF, and 
the term
$
\pi_{\text{evid}}(\mathbf{D})
$
is the evidence that is the probability of observing the data,
\begin{equation}
\pi_{\text{evid}}(\mathbf{D}) = \int \pi_{\text{like}}(\mathbf{D}|{\boldsymbol{\theta}})\cdot \pi_{\text{prior}}({\boldsymbol{\theta}}) \; \text{d}\boldsymbol{\theta}.
\end{equation}
%
In Bayesian calibration (\ref{eq:Bayes_calib}), the prior PDF reflects our initial knowledge about the model parameters. 
According to Jaynes \cite{jaynes2003}, if only parameters' bounds are available, i.e., complete ignorance, then uniform distribution should be taken into account as the parameter prior.
The form of the likelihood PDF, $\pi_{\text{like}}(\mathbf{D}|{\boldsymbol{\theta}})$ in (\ref{eq:Bayes_calib}), represents  the statistical distributions of discrepancy between the model output $\mathbf{d}(\boldsymbol{\theta})$ and the observational data $\mathbf{D}$.
Let {$p_\zeta$} be a probability distribution to the total error due to {modeling error,} $\boldsymbol{\xi}(\boldsymbol{\theta})$, and data noise, $\boldsymbol{\eta}$. Under the additive noise assumption, the total error is described as 
${\boldsymbol{\zeta}} = \boldsymbol{\eta} + \boldsymbol{\xi}(\boldsymbol{\theta}) = \mathbf{D} - \mathbf{d}(\boldsymbol{\theta})$ (see e.g., \cite{faghihi2018fatigue,prudencio2015}).
We assume that the total error is a Gaussian random variable of zero mean,
$
{\boldsymbol{\zeta}} \sim \mathcal{N}(\mathbf{0}, \boldsymbol{\Gamma}_{\text{noise}}^{-1}), 
$
where $\boldsymbol{\Gamma}_{\text{noise}}$ is a covariance matrix \cite{kaipio2006}.
The likelihood function is the probability density function describing the total error and written as,
\begin{equation}
\pi_{\text{like}}(\mathbf{D}|{\boldsymbol{\theta}}) = {p_\zeta}(\mathbf{D} - \mathbf{d}(\boldsymbol{\theta})).
\end{equation}
To explicitly represent the likelihood function, consider each data point as a sample from a distribution, $\mathbf{D}_i^{(j)} \sim p(\mathbf{D})$ with $j=1,\cdots,N_D$ is the independent and identically distributed (i.i.d.) realizations and $i=1,\cdots,N_t$ is the data points. The model output corresponding to each data point is denoted by $\mathbf{d}_i (\boldsymbol{\theta})$.
Assuming $\boldsymbol{\Gamma}_{\text{noise}} = (\sigma _i)^2 \mathbf{I}$, 
the log-likelihood function is,
\begin{eqnarray}\label{eq:likelihood}
\ln(\pi_{\rm like}(\mathbf{D}|\boldsymbol{\theta})) &= & 	
\sum_{i=1}^{N_t} \sum_{j=1}^{N_D}
\left[
-\frac{1}{2} \ln(2\pi) - \ln(\sigma_{i}) - \frac{1}{2} \left(
\frac{\mathbf{d}_i(\boldsymbol{\theta}) - \mathbf{D}_i^{(j)} }{\sigma_i}
\right)^2\right]. \nonumber\\
	& = &
\frac{1}{2} \sum_{j=1}^{N_D} \left( \mathbf{d}(\boldsymbol{\theta}) - \mathbf{D}_i^{(j)} \right)^T
\boldsymbol{\Gamma}_{\text{noise}}^{-1}
\left( \mathbf{d}(\boldsymbol{\theta}) - \mathbf{D}_i^{(j)} \right) + \mathrm{const}.
\end{eqnarray}

%

\subsubsection{Solution of Bayesian calibration}
For Bayesian model calibration (\ref{eq:Bayes_calib}), one requires to compute the posterior distribution as the solution of the statistical inverse problem, given the parameters' priors and the likelihood PDF. 
Markov Chain Monte Carlo (MCMC) sampling methods are employed in standard practice to characterize the posterior distribution. 
The MCMC solution of the Bayesian inference problem is computationally expensive due to the requirement of a large number of sequential model evaluations to explore possibly high-dimensional posterior distribution.
The Bayesian calibration of the SGP model, in which the model output is obtained by solving a highly nonlinear system of partial differential equations, requires parallelization of the MCMC and efficient use of computing resources. 

Metropolis-Hastings (MH) algorithm \cite{metropolis1953,hastings1970} is a commonly used class of MCMC sampling methods. 
This algorithm specifies an initial value $\boldsymbol{\theta}^{(0)}$ for the parameter $\boldsymbol{\theta}$, and at $l$-th iterations a candidate $\boldsymbol{\theta}^*$ is sampled from a proposal distribution $q(\cdot|\cdot)$. 
The most commonly used proposal density is a Gaussian distribution with fixed covariance and the mean, centered
at the value of the current state of the chain (random walk).
The $(l+1)$ step in the chain is,
\begin{equation}
\boldsymbol{\theta}^{(l+1)} = \left\{\begin{matrix}
\boldsymbol{\theta}^*  & {\rm with \; probability}& {\rm min} \{1,\alpha(\boldsymbol{\theta}^*,\boldsymbol{\theta}^{(l)}) \} \\ 
\boldsymbol{\theta}^{(l)} & {\rm with \; probability}& 1 - {\rm min}\{1,\alpha(\boldsymbol{\theta}^*,\boldsymbol{\theta}^{(l)}) \} 
\end{matrix}\right.
\end{equation}
where $\alpha$ is \textit{acceptance ratio},
\begin{equation}
\alpha(\boldsymbol{\theta}^*,\boldsymbol{\theta}^{(l)}) = 
\frac{\pi(\boldsymbol{\theta}^*)q(\boldsymbol{\theta}^{(l)}|\boldsymbol{\theta}^*)}
{\pi(\boldsymbol{\theta}^{(l)})q(\boldsymbol{\theta}^*|\boldsymbol{\theta}^{(l)})},
\end{equation}
where $\pi(\boldsymbol{\theta})$ being the posterior density, $\pi_{\rm post}(\boldsymbol{\theta})$.
For more details of MH and other MCMC algorithms, see, e.g., \cite{kaipio2006}.

{
In the Bayesian calibration results presented in section \ref{sec:dd-sgp}, we use an improved MH algorithm, known as Delayed Rejection Adaptive Metropolis (DRAM) \cite{haario2006dram}.
The MH sampler with Gaussian proposal distribution might lead to poor sampling if the proposal variance is too high. DRAM overcomes this deficiency by testing a series of back-up samples with smaller proposal variance before rejecting a candidate sample. If one of the back-ups is accepted, the MH algorithm continues, and if they are all rejected, the sampler rejects the candidate.
}

\subsubsection{Estimates and metrics}

Let us introduce the measures which be used in Section \ref{sec:results}
to interpret the computational results.
A point estimate representing the posterior PDF, $\pi_{\text{post}}({\boldsymbol{\theta}}|\mathbf{D})$, is the Maximum A Posteriori (MAP) point defined as,
\begin{equation}\label{eq:map}
\boldsymbol{\theta}^{\text{MAP}} = \underset{\boldsymbol{\theta}}{\text{argmax}} \; \pi_{\text{post}}(\boldsymbol{\theta}|\mathbf{D}).
\end{equation} 
The MAP can be computed by solving the deterministic inverse problem in (\ref{eq:map}) or approximated from the samples of the posterior distributions obtained from an MCMC algorithm.

To measure how well the observational data inform each model parameter,
we propose a measure based on the normalized variance of the posterior of each parameter $\theta_k, k=1,2,\cdots,K$ as
\begin{equation}\label{eq:info}
\mathcal{I} ({\theta}_k) = \frac
{\mathbb{V}\left(\pi_{\text{post}}(\theta_k|\mathbf{D})\right)}
{\mathbb{V}\left(\pi_{\text{prior}}(\theta_k)\right)}.
\end{equation} 
The $\mathbb{V}\left(\pi_{\text{post}}(\theta_k|\mathbf{D})\right)$ is the variance of the MCMC samples of the parameter posterior while the close form of the prior variance, $\mathbb{V}\left(\pi_{\text{prior}}(\theta_k)\right)$, is available analytically if the priors are the standard distributions, e.g., uniform, Gaussian.
The measure $\mathcal{I} ({\theta}_k)$ is the degree with which the model parameter ${\theta}_k$ is updated (learned from data) during parameter inference. 
The intuition behind the expression (\ref{eq:info}) is that a small $\mathcal{I}({\theta}_k)$ implies that the parameter variance is decreased significantly, from the prior to the posterior, and hence that parameter is well-informed by the data.
In the limit of $\mathcal{I}({\theta}_k) \rightarrow 1$, the posterior is identical to prior, and the model parameter is not learned from the data. 
Moreover, $\mathcal{I}({\theta}_k) \rightarrow 0$ indicates that the posterior is nearly delta function, and thus a high level of confidence in the parameter is gained through the inference.

Additionally, we make use of a metric to access the quality of the calibrated model in simulating observational data and predicting the QoI.
Let $\Pi^D(Q)$ and $\Pi^d(Q)$ be the cumulative distribution functions of the QoI, obtained from the model, $Q^d$, and from the data, $Q^D$.
The measure indicating the discrepancies among the model and data is given as \cite{odenbabuska2017},
\begin{equation}\label{eq:error}
\mathcal{E} = 
\frac{\int_{-\infty}^{\infty} | \Pi^D(\xi) - \Pi^d(\xi)| d\xi}{\mathbb{E}(Q^D)},
\end{equation}
where $\mathbb{E}(Q^D)$ is the mean of the QoI from observational data.

\subsection{Model prediction under uncertainty}\label{sec:forward}

Once the model parameters are calibrated using observational data, the computational model can be employed for predicting the quantity of interests (QoIs).
As indicated, Bayesian calibration consists of the solving the statistical inverse problem to obtain the parameter posterior $ \pi_{\text{post}}({\boldsymbol{\theta}}|\mathbf{D}) $. 
To assess computational prediction reliability, the uncertainty in parameters must be propagated through the model solution, resulting in the QoIs being random variables. To this end, the computational prediction is performed by solving the statistical forward problem.
{
The Monte Carlo method is frequently employed for forward uncertainty propagation. It involves drawing samples according to the parameter posteriors and evaluating the computational model outputs based on these samples. 
}

\section{Results}\label{sec:results}

In this section, we present the development of a sequential multiscale model of plasticity in small-volume materials in the presence of uncertainty. 
We begin by presenting the MDDP simulations of micro-pillar under compression, conducted over a wide range of sizes and initial dislocation contents. 
To investigate the SGP model prediction (described in Section \ref{sec:sgp}) in response to uncertain model parameters, we conduct the variance-based global sensitivity analyses.
Guided by the parameter sensitivity, we construct the discrete-continuum multiscale model by Bayesian calibration of the SGP using the MDDP simulations.
Finally, we investigate the multiscale model's reliability in predicting the micro-pillar size effect response considering different training (calibration) and tests (prediction) data sets of the MDDP simulations.

\subsection{MDDP simulations of micro-pillars}\label{sec:mddp_data}

We make use of the MDDP simulations of micro-pillars conducted by Shao et al. \cite{shao2014}, as synthetic data to inform the SGP model parameters while quantifying the uncertainty of the SGP model prediction of the size effect responses. 
The simulations consist of {uniaxial compression of micro-pillars with the aspect ratio of 1$\times$5$\times$1 with the height $L$ (along $y$ axis in Figure \ref{fig:pillar})} ranges from 200 nm to 1000 nm. The boundary conditions of the 3D micro-pillars are set as fixed displacement at one end, a prescribed displacement with a constant strain rate of $\dot{\epsilon} =-1$ s$^{-1}$ at the other end. 
To assess the microstructural randomness's effect on the size-dependent plasticity responses of micro-pillars, the MDDP simulations are conducted for different initial distributions of existing full and half Frank-Read sources along with initial dislocation densities ranging between $1.0 \mu m^{-2}$ to $100 \mu m^{-2}$.  
For each case of density-size combination, five different spatial distributions of dislocations are taken into account, including evenly distributed and concentrated dislocations in a portion of the domain length. 
{Other material parameters of the MDDP, such as elastic properties and dislocation mobility, are assumed to be the same for all the micro-pillars.}
Consequently, the MDDP simulations characterize the (macroscopic) size effect of micro-pillars' plastic deformation under different (microscopic) initial density and distribution of dislocations. More details of the MDDP simulations are provided in \cite{shao2014}.

\begin{figure}[!htb]
\centering
\includegraphics[trim = 0mm 0mm 0mm 0mm, clip, width=.48\textwidth]{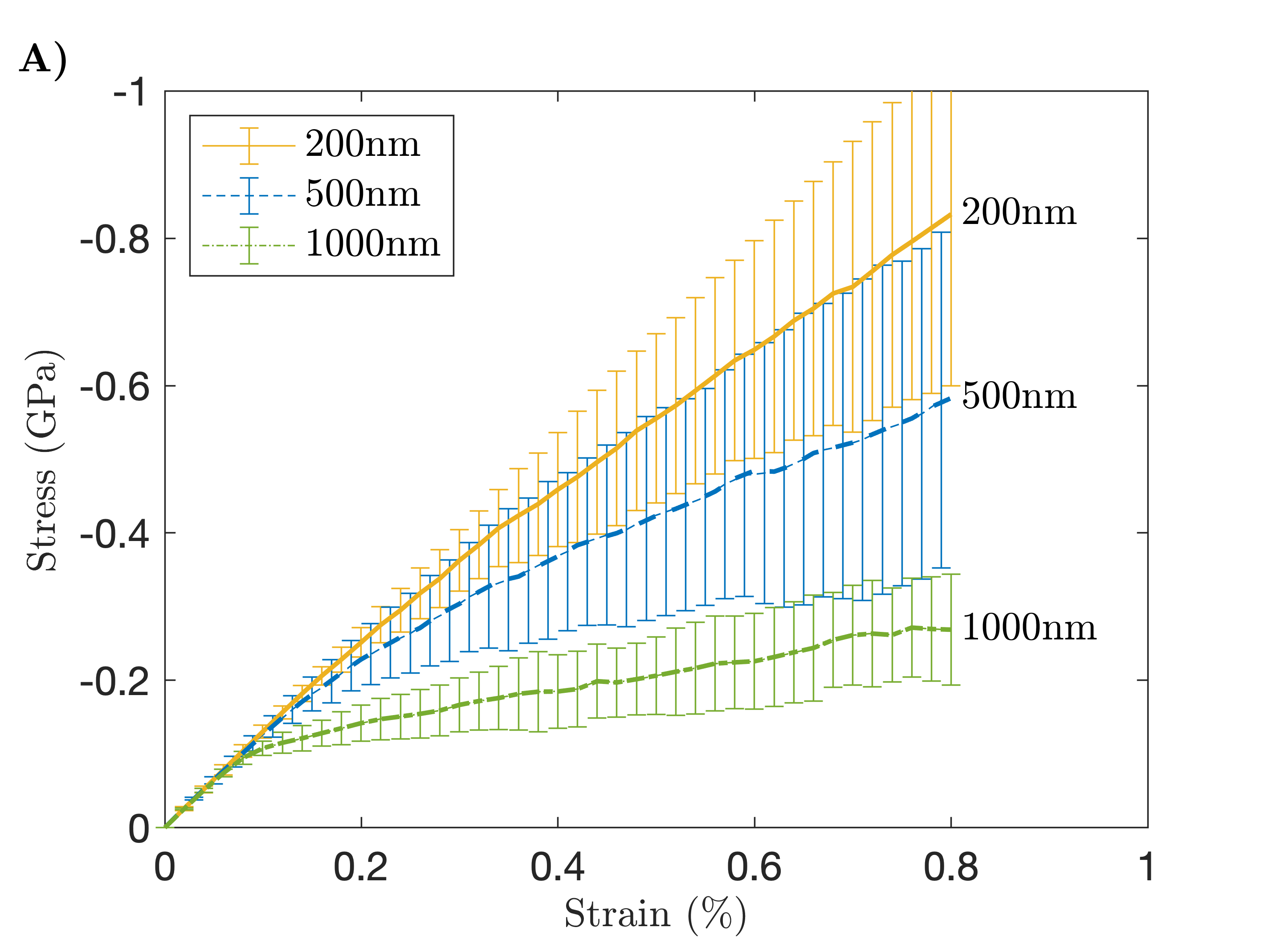}
~
\includegraphics[trim = 0mm 0mm 0mm 0mm, clip, width=.48\textwidth]{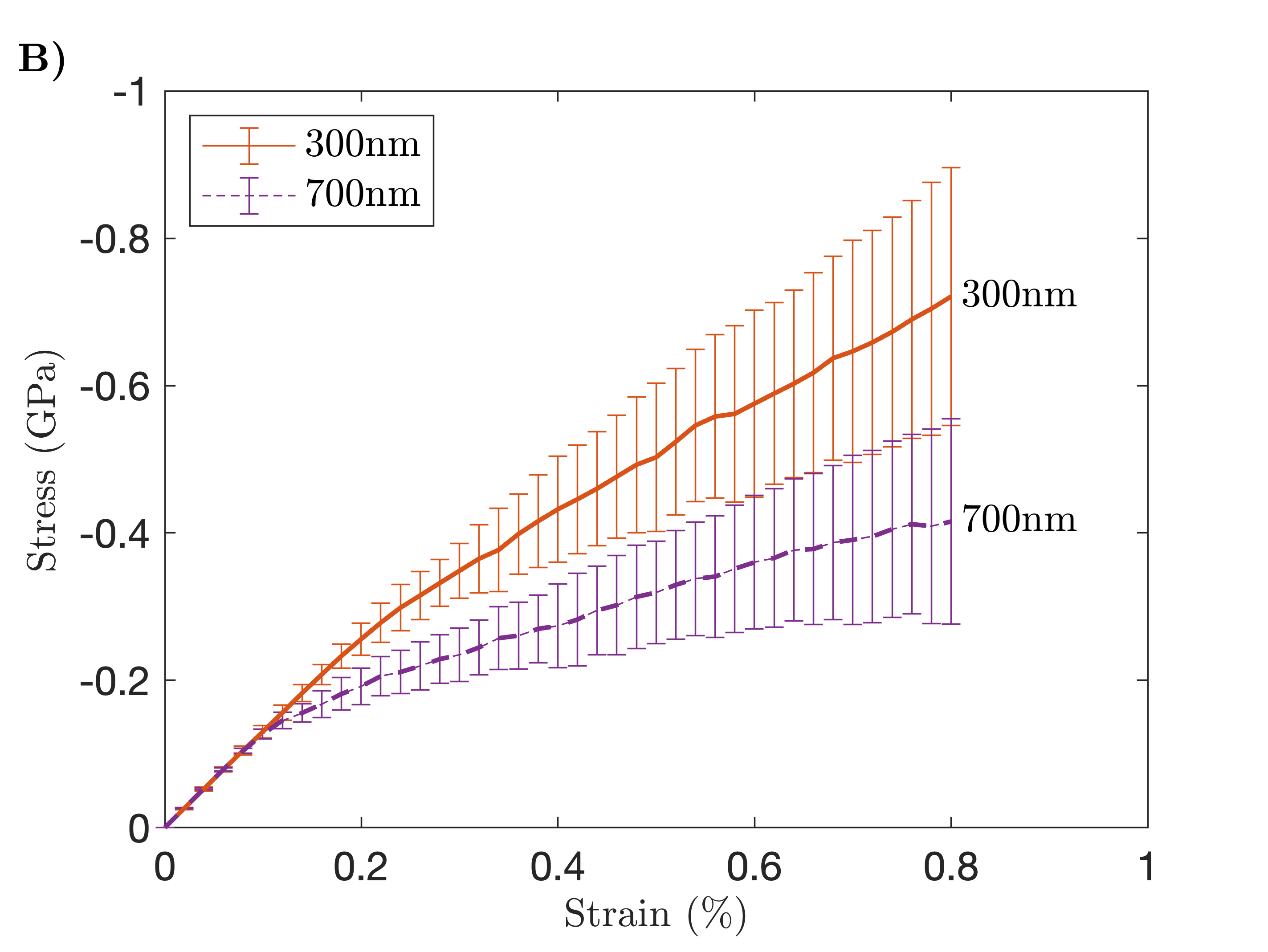}
\vspace{-0.05in}
\caption{
{Stress-strain synthetic data generated by MDDP simulations of micro-pillar compressions with the sizes 
$L = $ 200 nm, 300 nm, 500 nm, 700 nm, and 1000 nm \cite{shao2014}. 
The standard deviation error bars for each size represents the stochasticity of MDDP simulations estimated using five realizations, different initial dislocation density ($3.16 \mu m^{-2}$ and $30.16 \mu m^{-2}$), and spatial distributions (uniformly distributed and concentrated in some regions of the domain). The average variances of the data are 16.64\%, 14.96\%, 26.40\%, 18.27\%, and 21.78\% for the 200 nm to 1000 nm micro-pillars, respectively.  For better presentation, the error bars are shown only every 50 data points. }
}
\label{fig:data}
\vspace{-0.1in}
\end{figure}

To assess the reliability of the MDDP to SGP multiscale model prediction, special attention must be paid to characterizing the various sources of uncertainty in micro- and macro-scales. One major contributor to the uncertainty is that the continuum SGP model filters out the detailed microstructural evolutions simulated by MDDP.
As shown in Section \ref{sec:forward}, the SGP model is capable of capturing the size dependency of plastic response in micro-pillars. However, this continuum model fails to account for the microscopic effect of initial dislocations' density and their heterogeneity on the plastic deformation. 
An additional source of uncertainty is the inherent randomness of dislocation evolutions in the MDDP approach due to the stochastic algorithms that govern the dislocation interactions (data noise).
{
Further uncertainty stems from the finite element approximation of the SGP model and the use of a 1D model for simulating micro-pillars. The 1D approximation results in an additional modeling error since it does not account for the free boundaries on the stress-strain responses.}
In developing the multiscale scale model, we characterize these uncertainties through a statistical representation of the microstructural randomness. In other words, we view the macroscopic size effect in the stress-strain responses of the MDDP simulations as the observational data, while the microstructural effects of the initial dislocation configuration on the size effect are considered as epistemic data uncertainty. We then calibrate the SGP against these synthetic data furnished by MDDP simulations while accounting for both continuum model inadequacy and the noise in discrete simulations.
Figure \ref{fig:data} shows the synthetic data used for the calibration of the SGP model. The data are generated by the MDDP simulations of micro-pillar with the sizes $200$ nm, $300$ nm, $500$ nm, $700$ nm, and $1000$ nm. The error bars in Figure \ref{fig:data} represent the uncertainty due to different initial dislocation density, the spatial distribution of dislocations, and five realizations of the MDDP simulations for each micro-pillar size. 
The mean of the synthetic data clearly shows the size effect phenomena, in which smaller micro-pillar express higher flow stresses and hardening rates. 
However, due to the randomness in initial dislocation contents
there is considerable uncertainty in the stress-strain results with a 20\% of the average variance in stress.
The significant uncertainty in the synthetic data demands rigorous uncertainty treatment methods for learning the SGP model from the MDDP data and determining the level of confidence in the multiscale model prediction of plasticity responses. 
%

\subsection{Numerical analysis of the SGP}\label{sec:forward}
\begin{figure}[!htb]
\centering
\includegraphics[trim=0.0in 0.0in 0.0in 0.0in, clip, width=.25\textwidth]{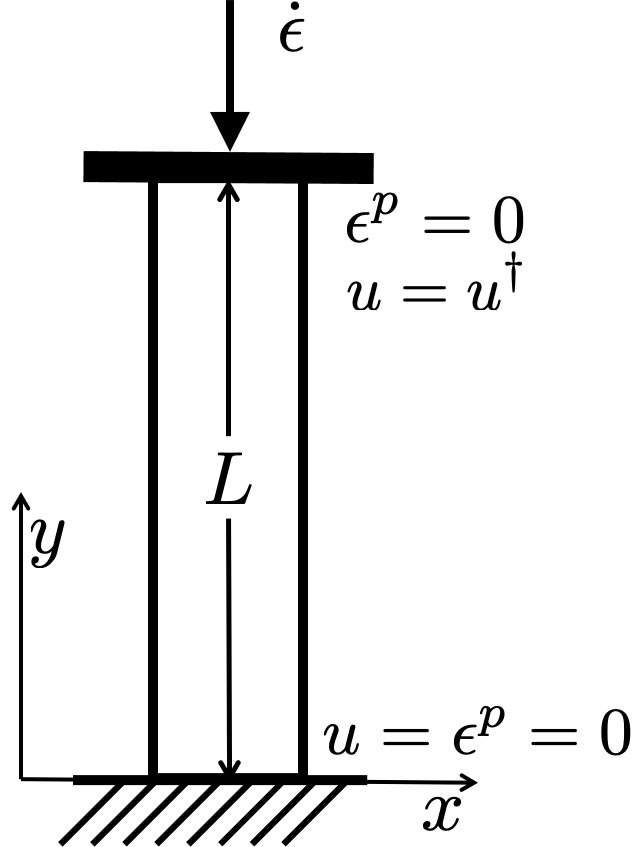}
\vspace{-0.05in}
\caption{
{Schematic illustration of the micro-pillars with size (height) $L$ under compression along with the
boundary conditions imposed to the top and bottom of the 1D SGP simulation domain.}
}
\label{fig:pillar}
\vspace{-0.1in}
\end{figure}

The 1D finite element solution of the SGP model (Section \ref{sec:fem}) is implemented in an open-source computing platform, FEniCS \cite{fenics}. 
In all SGP simulations, micro-pillars are modeled by discretizing the domain by 30 elements. 
{Consistent with the MDDP simulations of micro-pillars (section \ref{sec:mddp_data}),} we consider the macroscopic boundary conditions of the SGP as $u{(y=0)}=0$ on one end of the domain, and the other end {$y=L$} is subjected to incremental displacements with a constant strain rate of $\dot{\epsilon} =-1$ s$^{-1}$. Additionally, micro-clamped conditions, $\epsilon^p=0$, are imposed at the boundaries{, to mimic the impenetrable boundaries used in MDDP simulations that block the dislocation movement at the top and bottom of the domain.} 
{
Figure \ref{fig:pillar} illustrates the micro-pillar compressions and the boundary conditions imposed to the top and bottom of the 1D SGP simulation domain.}
Following a convergence study of the Newton solver for various combinations of the parameter values, we use the time increments of $\Delta t = 5.0\times 10^{-5}$ s for applying the incremental displacement. 
Since the MDDP simulations do not account for the rate dependency, the visco-plastic parameters of the SGP model are fixed, $m=0$ and $q=1$.

\begin{figure}[!htb]
\centering
\includegraphics[trim = 0mm 0mm 0mm 0mm, clip, width=.48\textwidth]{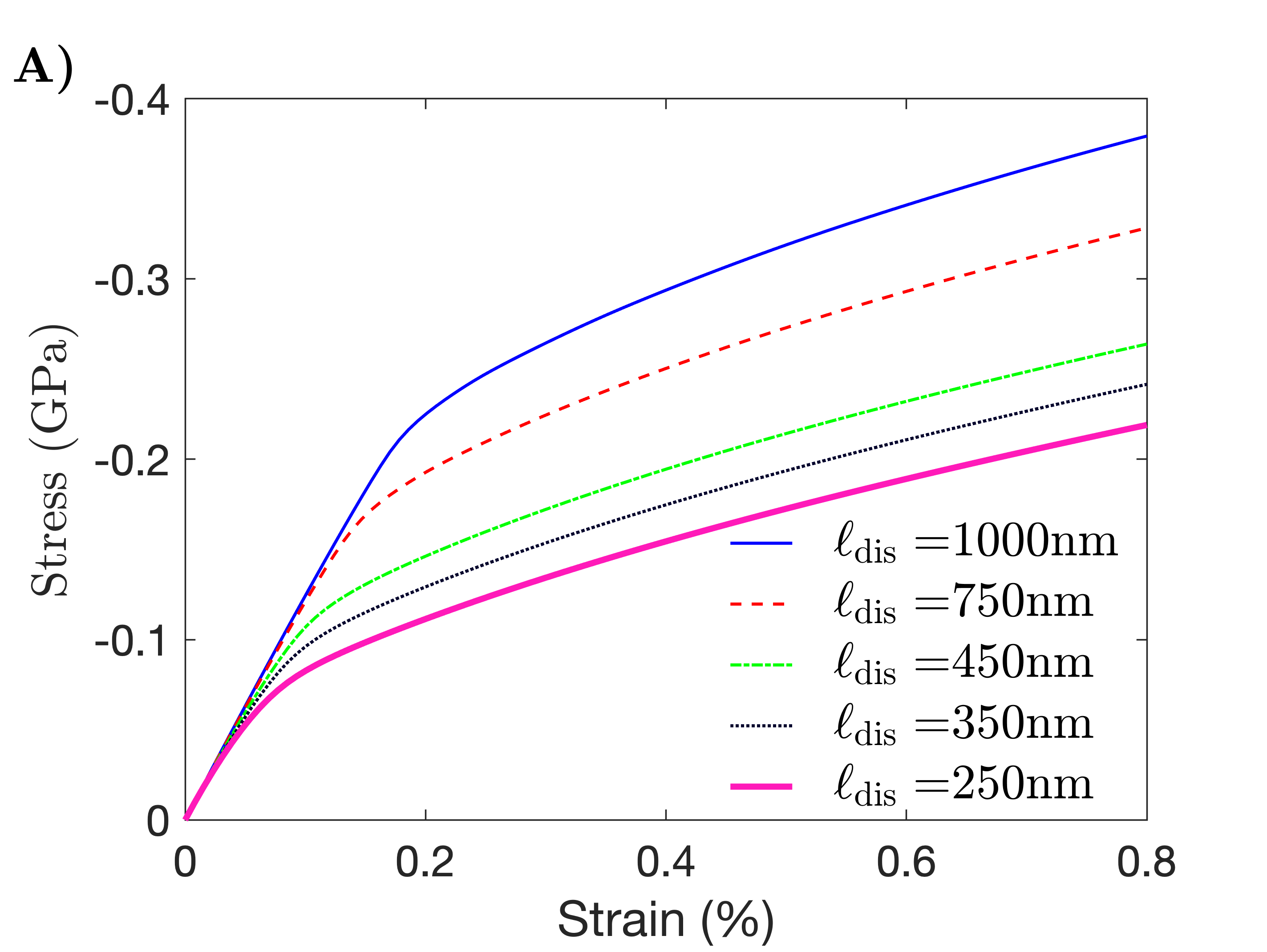}
~
\includegraphics[trim = 0mm 0mm 0mm 0mm, clip, width=.48\textwidth]{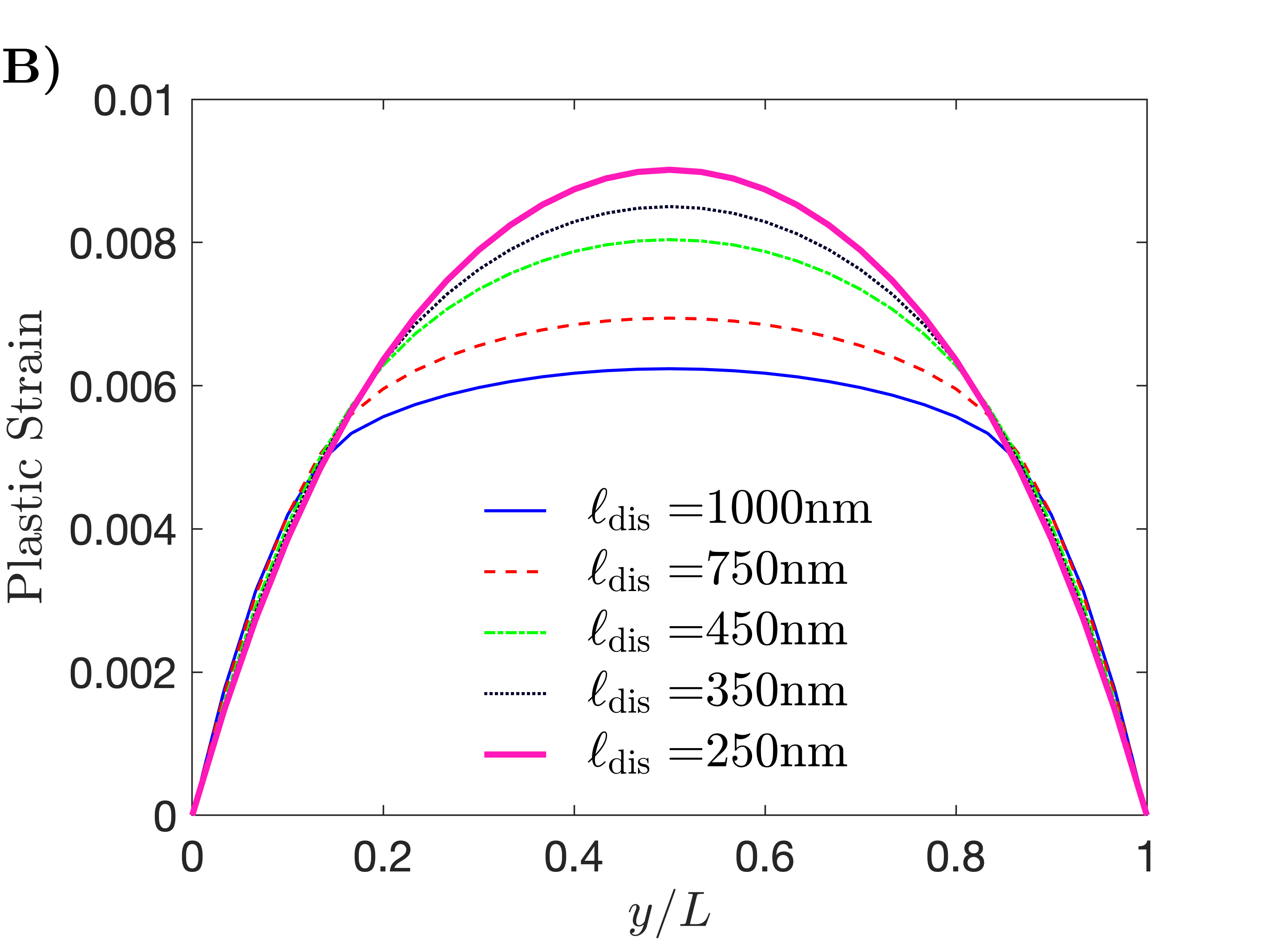}
~
\includegraphics[trim = 0mm 0mm 0mm 0mm, clip, width=.48\textwidth]{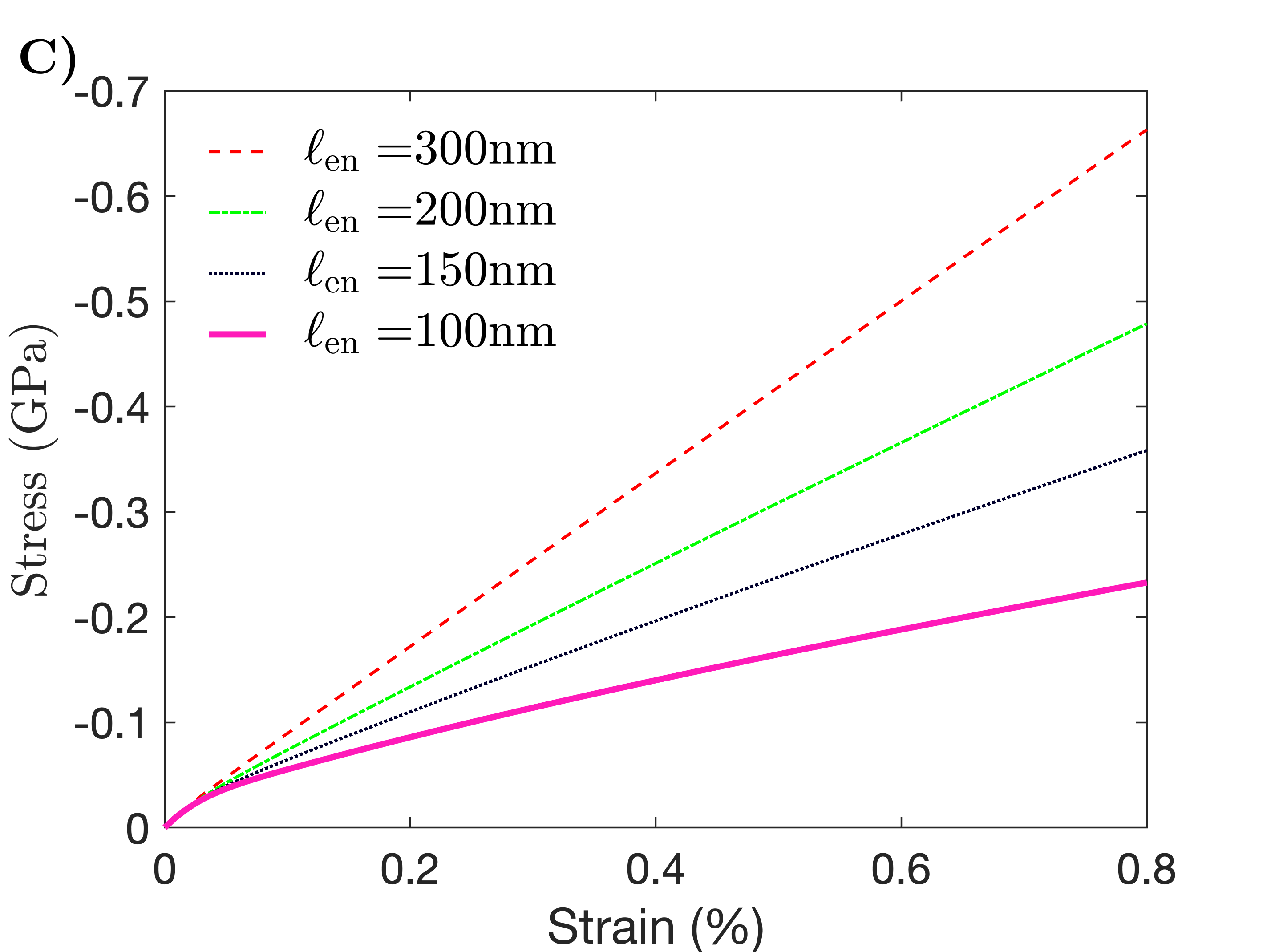}
~
\includegraphics[trim = 0mm 0mm 0mm 0mm, clip, width=.48\textwidth]{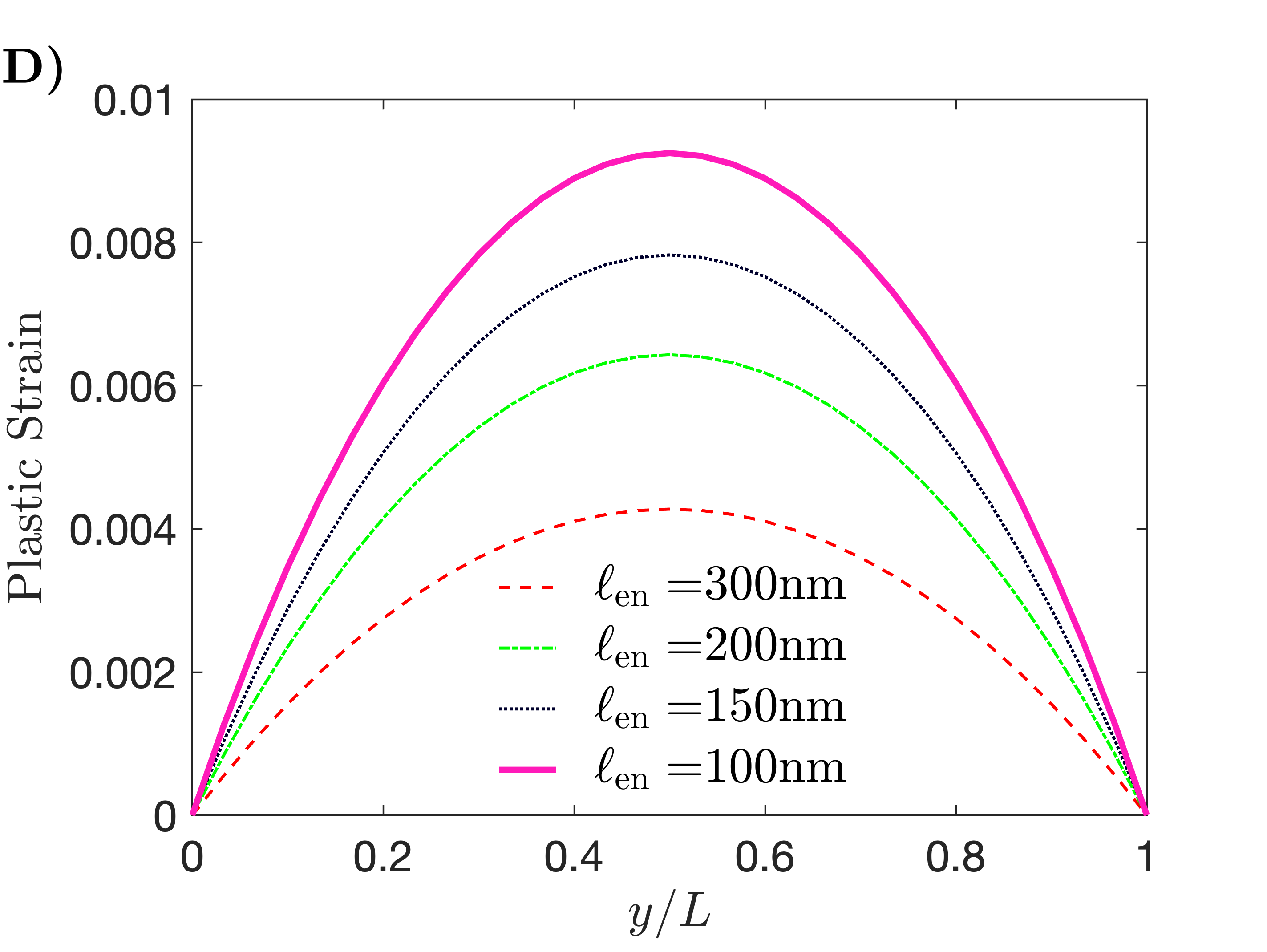}
\vspace{-0.15in}
\caption{
{Numerical experiments of the stress-strain response of the micro-pillar with the size $L=500$nm using the strain-gradient plasticity model.
(A and C) Stress-strain plot during compression,
and
(B and D) spatial distribution of plastic strain across the micro-pillar at a macroscopic applied uniaxial strain of 0.8\%.
The size effect is shown by
(A and B) the variation of the dissipative length scale $\ell_{\text{dis}}$ with the energetic length scale $\ell_{\text{en}} = 75$nm,
(C and D) the variation of the energetic length scale $\ell_{\text{en}}$ with the dissipative length scale $\ell_{\text{dis}} = 20$nm.
The other parameters of the model are: $E = 128.44$GPa, $Y = 0.047$GPa, $h = 0.062$GPa, and $r = 298.42$.
}}
\label{fig:sgp}
\vspace{-0.1in}
\end{figure}

A set of numerical experiments are conducted to study the size effect responses of the SGP model. 
{The analyses consist of micro-pillars with the {height $L=500$nm} undergoing compression up to a macroscopic applied uniaxial strain of 0.8\%.
Figure \ref{fig:sgp} presents the finite element results of the stress-strain variations (panels A and C) and spatial distribution of the plastic strain across the micro-pillar at the strain of 0.8\% (panels B and D). }
The effect of size on stress-strain responses due to variation of $\ell_{\text{dis}}$ (Figure \ref{fig:sgp} A) {clearly shows that the} dissipative length scale affects the flow stress (onset of plasticity) while the hardening rate is the same in all the plots. 
Additionally, increasing $\ell_{\text{en}}$ results in more significant kinematic hardening as shown in Figure \ref{fig:sgp} (C). 
The plastic strain profile is shown in Figure \ref{fig:sgp}(C and D) indicates that discretizing the plastic strain and displacement by different finite element interpolation functions results in an accurate solution without a need for a high number of elements to resolve the incompatibility between the $u$ and $\epsilon^p$ solutions as in \cite{faghihi2012phd, faghihi2014jemt, anand2005}. 
The numerical experiments presented in these plots show that, for small values of the dissipative length scale, the development of a small thickness boundary layer with a sharp plastic strain gradient is observed at the vicinity of the boundaries. 
In summary, the numerical experiments presented in Figure \ref{fig:sgp} indicates that the SGP qualitatively represents the microscopic dislocation phenomena leading to macroscopic size effect phenomena, as observed in the microscale experiments, e.g., \cite{xiang2005, xiang2006, camposilvan2016}.
%

\subsection{Global sensitivity analysis  of the SGP}

To determine the effect of each model parameter contribution to the SGP model prediction, we perform global sensitivity analyses. Corresponding the MDDP simulations described in Section \ref{sec:mddp_data}, the parameter sensitivity scenarios consist of 
micro-pillars with the sizes of 200 nm to 1000 nm under compression from zero strain until the strain reaches $\epsilon =$ 0.8\%.
The target QoI for computational prediction is considered as the total strain energy of the micro-pillar,
\begin{equation}\label{eq:qoi}
Q = \int_0^{\epsilon=0.8\%} |T| \; d{\epsilon}.
\end{equation}
%

\begin{figure}[!htb]
\centering
\includegraphics[trim = 0mm 0mm 0mm 0mm, clip, width=.48\textwidth]{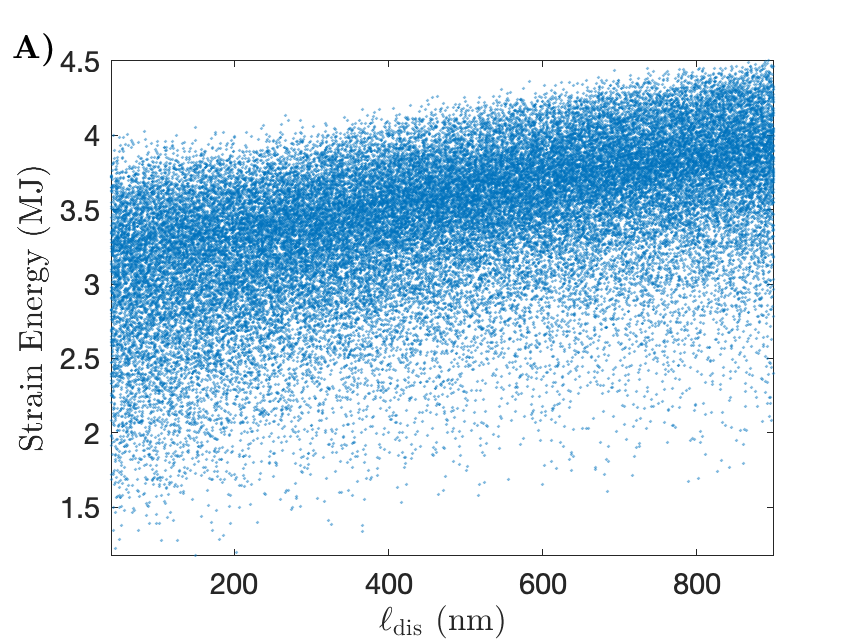}
~
\includegraphics[trim = 0mm 0mm 0mm 0mm, clip, width=.48\textwidth]{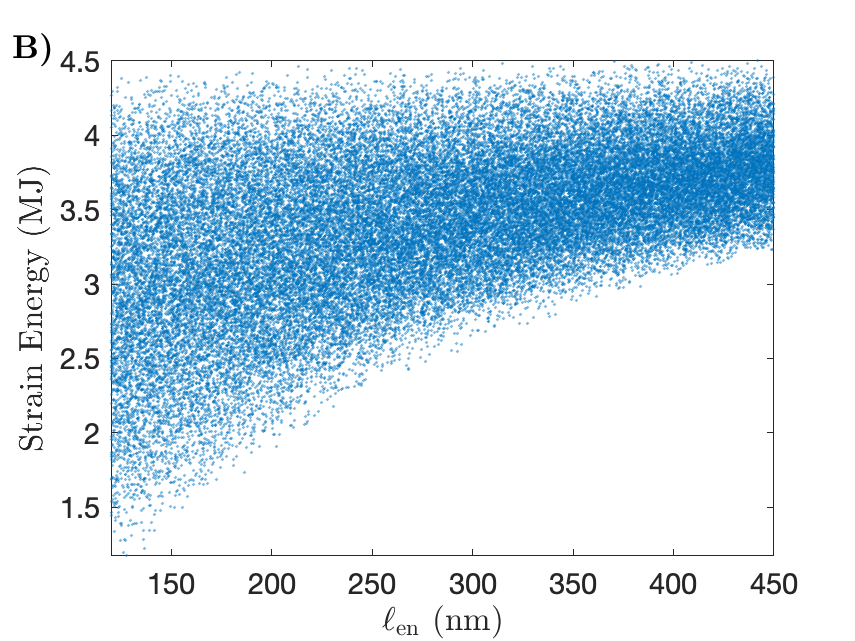}
~
\includegraphics[trim = 0mm 0mm 0mm 0mm, clip, width=.48\textwidth]{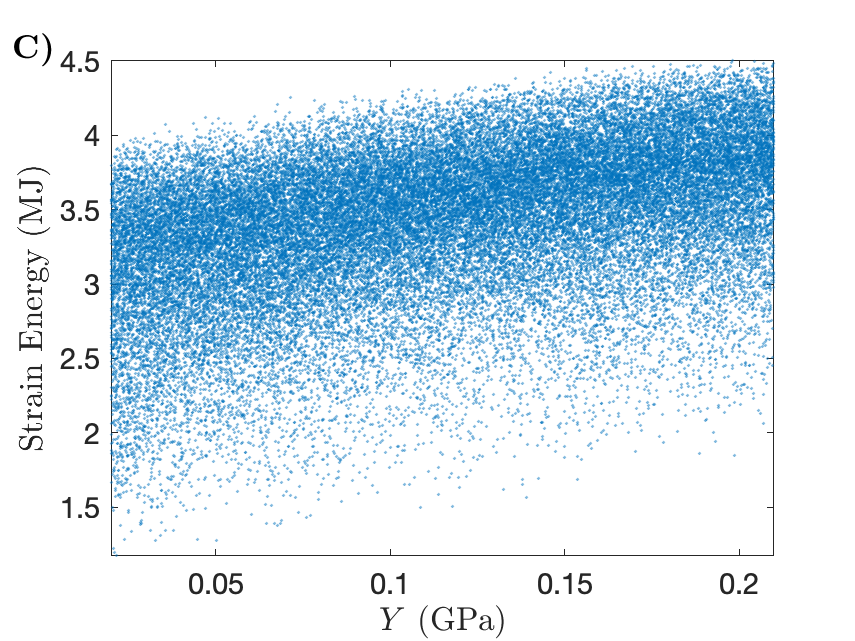}
~
\includegraphics[trim = 0mm 0mm 0mm 0mm, clip, width=.48\textwidth]{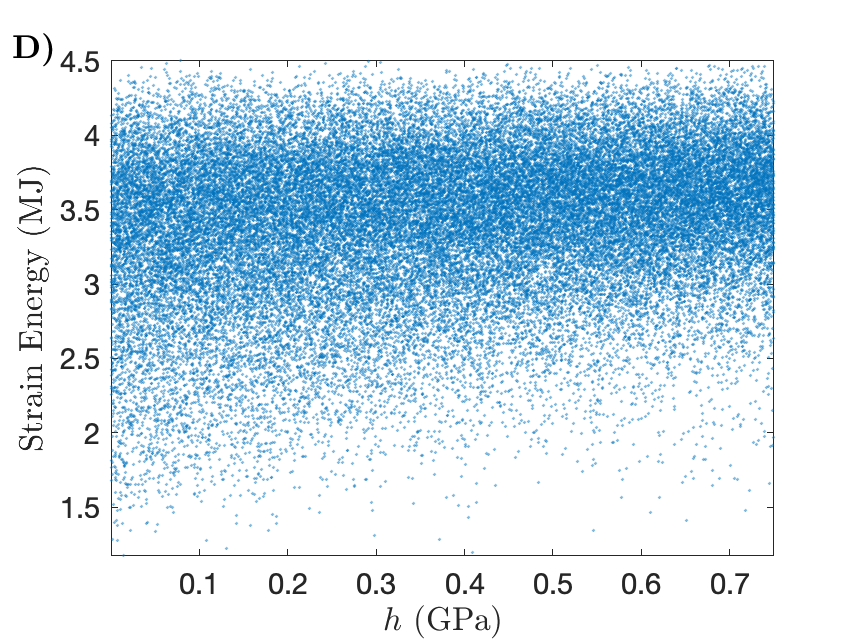}
~
\includegraphics[trim = 0mm 0mm 0mm 0mm, clip, width=.48\textwidth]{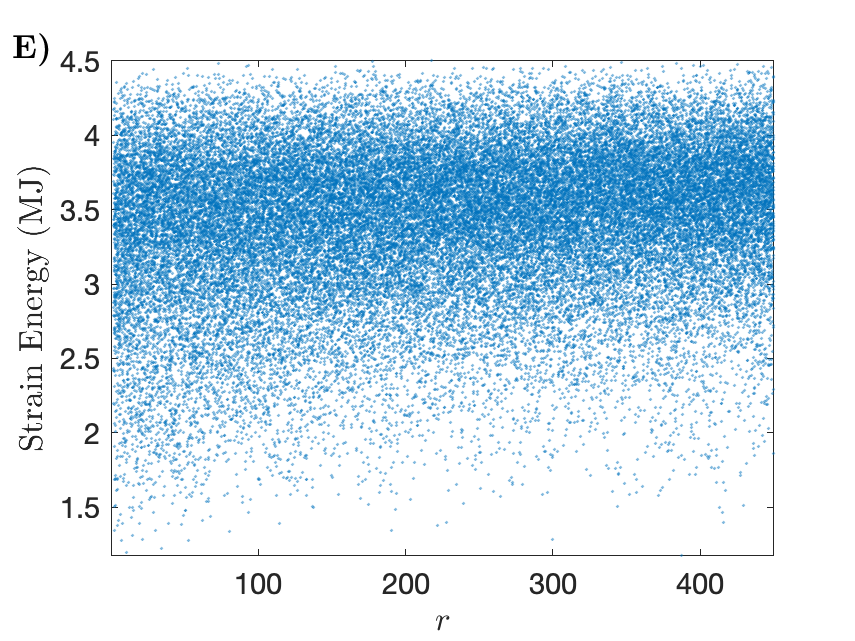}
~
\includegraphics[trim = 0mm 0mm 0mm 0mm, clip, width=.48\textwidth]{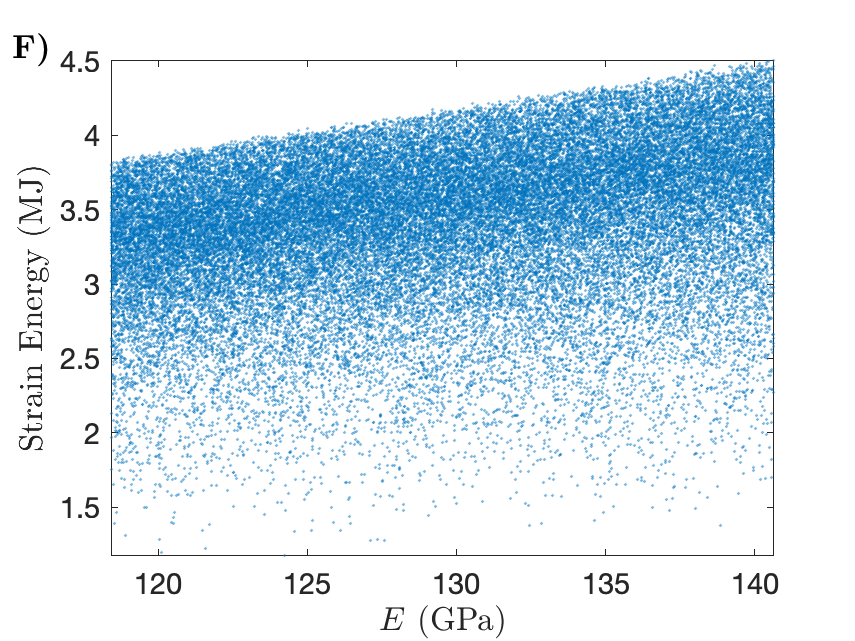}
\vspace{-0.15in}
\caption{
Scatter-plots of the SGP parameters for a micro-pillar with the size of 500nm under compression.
The QoI is the strain energy of the micropillars. The 50000 samples are drawn from the probability distributions of the uncertain parameters presented in Table \ref{table:param} using LHS. 
Distinct patterns observed in scatter-plots of $\ell_{\text{en}}$ and $\ell_{\text{dis}}$ indicate that they are the most important parameters of the model.
}
\label{fig:scatter}
\vspace{-0.1in}
\end{figure}

A direct visual indication of parameter sensitivity is scatter-plots \cite{saltelli2008}, consisting of clouds of the model outputs versus random variations in all the input parameters. The parameters with a significant impact on the model output are the ones with a distinct pattern \cite{saltelli2008} in the scatter-plot. The scatter-plots of the SGP are shown in Figure \ref {fig:scatter} for a micro-pillar with 500 nm of size. The uncertain model parameters, $\boldsymbol{\theta} = (\ell_{\text{dis}}, \ell_{\text{end}}, Y, h, r, E)$, are represented by the uniform probability distributions according to Table \ref{table:param}. The strain energy (\ref{eq:qoi}) is then computed for 50000 samples of the parameters according to their uncertainty range. In particular, the Latin Hypercube Sampling (LHS) method \cite{helton2003} is employed to distribute the samples evenly over the multi-dimensional parameter space.
It is seen from the scatter-plots of Figure \ref {fig:scatter} that the length scales $\ell_{\text{en}}$ and $\ell_{\text{dis}}$ are the most important contributors to the micro-pillar strain energy, as they are exhibiting distinct patterns of the clouds in the scatter-plots.

%

\begin{figure}[!htb]
\centering
\includegraphics[width=1.0\textwidth]{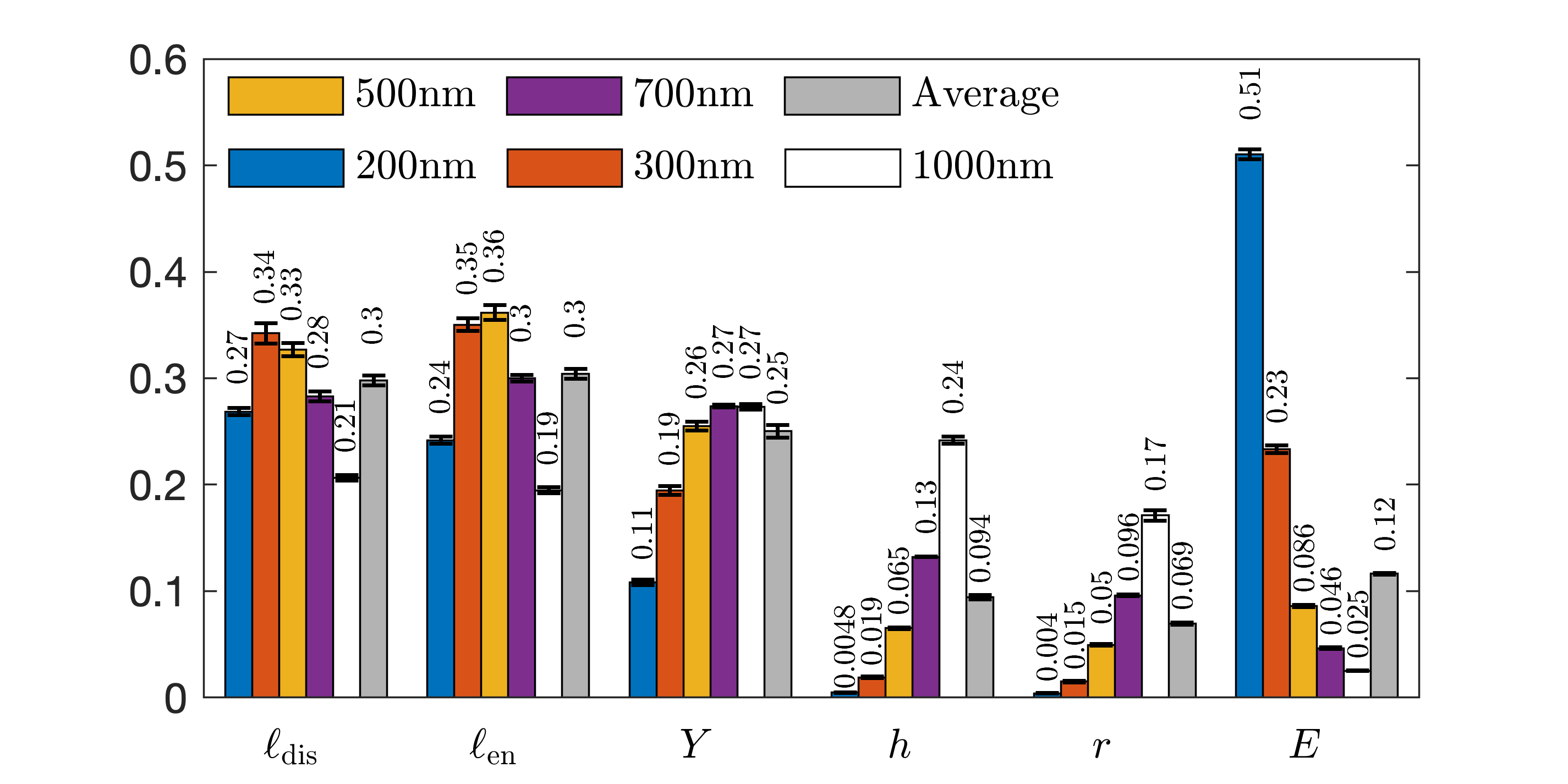}
\vspace{-0.25in}
\caption{
The results of variance-based global sensitivity analysis of the SGP model for the strain energy of different micro-pillar sizes and the average of the strain energies.
The total effect sensitivity indices $S_k, k=1,2, \cdots, 6 $ are estimated by (\ref{eq:STapprox}) and using $N$=10000 samples drawn from uncertain parameters according to the distributions in Table \ref{table:param}.
On average, the length scales $\ell_{\text{en}}$ and $\ell_{\text{dis}}$ are the most important parameters, while small values of the sensitivity indices for $h$ and $r$ indicate that
{the} variability of these parameters has insignificant contributions to the uncertainty in SGP model prediction.
}
\label{fig:sensitivity}
\vspace{-0.1in}
\end{figure}

A quantitative approach to rank the model parameter importance is the variance-based global sensitivity analyses described in Section \ref{sec:sensitivity}. We performed this sensitivity analysis on the SGP model and for the micro-pillars with the sizes of 200 nm to 1000 nm.
To this end, $N$=10000 samples are drawn for the parameters using Latin Hypercube sampling, and the total effect sensitivity indices $S_k, k=1,2, \cdots, K= 6 $ are computed using the estimator (\ref{eq:STapprox}), resulting in 80000 model evaluations. 
In addition to parameter sensitivity for each micro-pillar size, the average
total effect sensitivity indices are computed by considering the model output $Q$ as the mean of the strain energies over all the sizes.
To ensure sufficient samples are used, the Monte Carlo estimation of the indices are repeated four times.
The total sensitivity indices for each micro-pillar size and the average indices are shown in Figure \ref{fig:sensitivity}. The relatively small error bars (with an average variance of 2\%) indicate that an adequate number of samples is used to explore the six-dimensional parameter space.
%

The average total sensitivity indices in Figure \ref{fig:sensitivity} shows that, in the strain and size ranges of the MDDP simulations (Section \ref{sec:mddp_data}), length scales $\ell_{\text{en}}$ and $\ell_{\text{dis}}$ are the most contributors to the micro-pillar strain energy. These results are expected as the length scales control the size dependency of the flow stress and hardening rate, as shown in Figure \ref{fig:sgp}. 
Additionally, the isotropic hardening parameters $h$ and $r$ shows an insignificant impact on the model output and the SGP model prediction, judged by smaller values of the mean of the total sensitivity indices. 
However, Figure \ref{fig:sensitivity} indicates that the parameter sensitivity strongly depends on the micro-pillar size. 
The elastic modulus $E$ shows the most pronounced variation of sensitivity indices with size, in which a 181.31\% difference in $S$ is observed between the sizes of 200 nm to 1000 nm. An opposite trend is observed in the yield strength $Y$ and hardening parameters $h$ and $r$, where their impact on strain energy increases with micro-pillar size. 
These observations attribute to the size dependency of the stress-strain responses. That is, smaller micro-pillars show higher strength, and thus variation in elastic property significantly affects the strain energy. As the micro-pillar sizes increase, the impact of classical plasticity parameters $(Y, h, r)$ dominates the sensitivity of $E$. These responses agree with the parameter sensitivity of plasticity deformation in macroscale materials \cite{odenbabuska2017}.
Figure \ref{fig:sensitivity} also shows non-monotonic variation of the sensitivity indices of the length scales, $\ell_{\text{en}}$ and $\ell_{\text{dis}}$, with the micro-pillar sizes, in which the value of $S$ is higher for the sizes of 300nm and 500nm. 
Such observations can be interpreted by the effect of the domain size on the plastic strain gradient development in the small volume materials. 
For larger micro-pillars, the effect of plastic strain gradient diminishes, such that on the limit of macroscale materials, the length scales vanish, corresponding to the classical plasticity theory. On the other hand, the micro-pillar boundary layer development is restricted by the domain size for very small micro-pillars. This limitation leads to a lower impact of the length parameters on the stress-strain responses and strain-energy, once the material size is very small, i.e., micro-pillars with size $\leq 200$ nm.

\begin{table}[!ht]
\centering
\caption{
SGP model parameters.
Probability distributions of parameters considered for the sensitivity analyses,
and priors for the Bayesian model calibration. 
The $\mathcal{U}(\cdot,\cdot)$ indicates the uniform probability distribution.
}
\begin{tabular}{c|c|c}
\hline    
\textbf{Parameter}  
& \textbf{Physical meaning}      
& \textbf{Probability distributions} \\\hline
$\ell_{\text{dis}}$ & dissipative length scale             & $\mathcal{U}(40 , 900)$ nm \\
$\ell_{\text{en}} $ & energetic length scale              & $\mathcal{U}(120 , 450)$ nm  \\
$Y$                   
& yield strength                                         & $\mathcal{U}(0.02 , 0.21)$ GPa  \\
$h$                   
& isotropic hardening parameter   & $\mathcal{U}(0.00001 , 0.75)$ GPa \\ 
$r$                   
& isotropic hardening exponent      & $\mathcal{U}(0.2 , 450)$  \\ 
$E$                   
& Elastic modulus                                      & $\mathcal{U}(118.43, 140.64)$ GPa  \\ 
$m$                   
& strain rate power                                    &  fixed at the value $m=0$ \\ 
$q$                     
& strain rate coefficient                             &   fixed at the value $q=1$ s$^{-1}$  \\ \hline
\end{tabular}
\label{table:param}
\end{table}

\subsection{Developing MDDP to SGP multiscale model}\label{sec:dd-sgp}


After studying each parameter's effect on the SGP model output's uncertainty using sensitivity analyses, {we develop the sequential discrete-continuum multiscale model, by Bayesian calibration of the SGP model parameters using the MDDP simulation data (Figure \ref{fig:data}).}
Our model calibration process consists of splitting the synthetic data of different micro-pillar sizes (Figure \ref{fig:data}) into 
(i) \textit{training sets:} the MDDP scenarios (micro-pillar sizes) that are used to inform the SGP through Bayesian calibration;
(ii) \textit{testing sets:} the scenarios that are not included in the calibration process, i.e., unseen data sets, and used to challenge the calibrated SGP model's predictive capability.
In other words, we make use of Bayesian inference to determine the SGP parameters that represent the training data sets and compare the calibrated SGP model results with the testing data to investigate the model's validity in predicting the size effect responses outside the calibration data. 
{Here we identify two cases, with different combinations of training and testing sets of the MDDP simulations (Table \ref{tabel:scenarios}). In Case I, the calibration data $\mathbf{D}$ is the stress-strain results of MDDP obtained from the micro-pillars with the sizes of 300 nm, 500 nm, 700 nm, and 1000 nm and the testing set is the 200 nm micropillar. In Case II, the testing set is the MDDP results of the 1000 nm micro-pillar, and the stress-strain of other micro-pillars are the training sets.}

\begin{table}[!ht]
\centering
\caption{
{Mean and standard deviation of the error in QoI (strain energy) between the MDDP and SGP simulations.
Each case considers different training/testing sets of synthetic data furnished by MDDP for Bayesian calibration of the SGP.}
}
\begin{tabular}{c|c|c|cc}
\hline
micro-pillar & \multicolumn{2}{c|}{\textbf{Case I}} & \multicolumn{2}{c}{\textbf{Case II}}  \\ \cline{2-5} 
size     	& sets      & $\mathcal{E}$     & \multicolumn{1}{c|}{sets}  & $\mathcal{E}$ \\ \hline
1000 nm   & training     & 0.049$\pm$0.21 & \multicolumn{1}{c|}{testing}  &  0.31$\pm$0.19  \\
700 nm    & training     &  0.050$\pm$0.20 & \multicolumn{1}{c|}{training} &  0.03$\pm$0.19  \\
500 nm    & training     &  0.056$\pm$0.30 & \multicolumn{1}{c|}{training} &  0.07$\pm$0.30  \\
300 nm    & training     &  0.053$\pm$0.17 & \multicolumn{1}{c|}{training} &  0.008$\pm$0.16  \\
200 nm    & testing      &  0.082$\pm$0.19 & \multicolumn{1}{c|}{training} &  0.04$\pm$0.19  \\ \hline
\end{tabular}
\label{tabel:scenarios}
\end{table}
%


To conduct the Bayesian calibration, we make use of the likelihood function in (\ref{eq:likelihood}), in which the model output $\mathbf{d}(\boldsymbol{\theta})$ is the stress computed from the SGP model.
As indicated in the introduction section, the parameter calibration of SGP has not been widely conducted in the literature and our prior knowledge about these parameters is limited.
We thus assumed the priors being uniform distributions and performed two deterministic calibrations of the SGP model using the observational data's lower and upper bounds to determine the priors.
First, the minimum and maximum stress values at each strain level of the MDDP simulations are computed for the micro-pillar sizes of 1000 nm and 200 nm, respectively.
Then, the maximum and minimum values of the parameter priors are computed by calibrating the model against the synthetic data's upper and lower bound using the least-squares method. The resulting priors of the parameters are presented in Table \ref{table:param}.


The Bayesian analyses are conducted using the DRAM algorithm as implemented in a parallel object-oriented statistical library, Dakota (v. 6.12) \cite{ dakota2020}. 
For calibrating the SGP model, ten MCMC chains are used with chain lengths of 10000. Each chain was initialized from different parameter values within the priors' range to explore the parameter space better and check if they lead to similar results. To allow the Markov chain to get sufficiently close to the stationary distribution, we consider the initial $10\%$ of the chains as a burn-in period.


\subsubsection{{Case I calibration and prediction}}

\begin{figure}[!htb]
\centering
\includegraphics[trim=0in 0in 0in 0in, clip, width=1\textwidth]{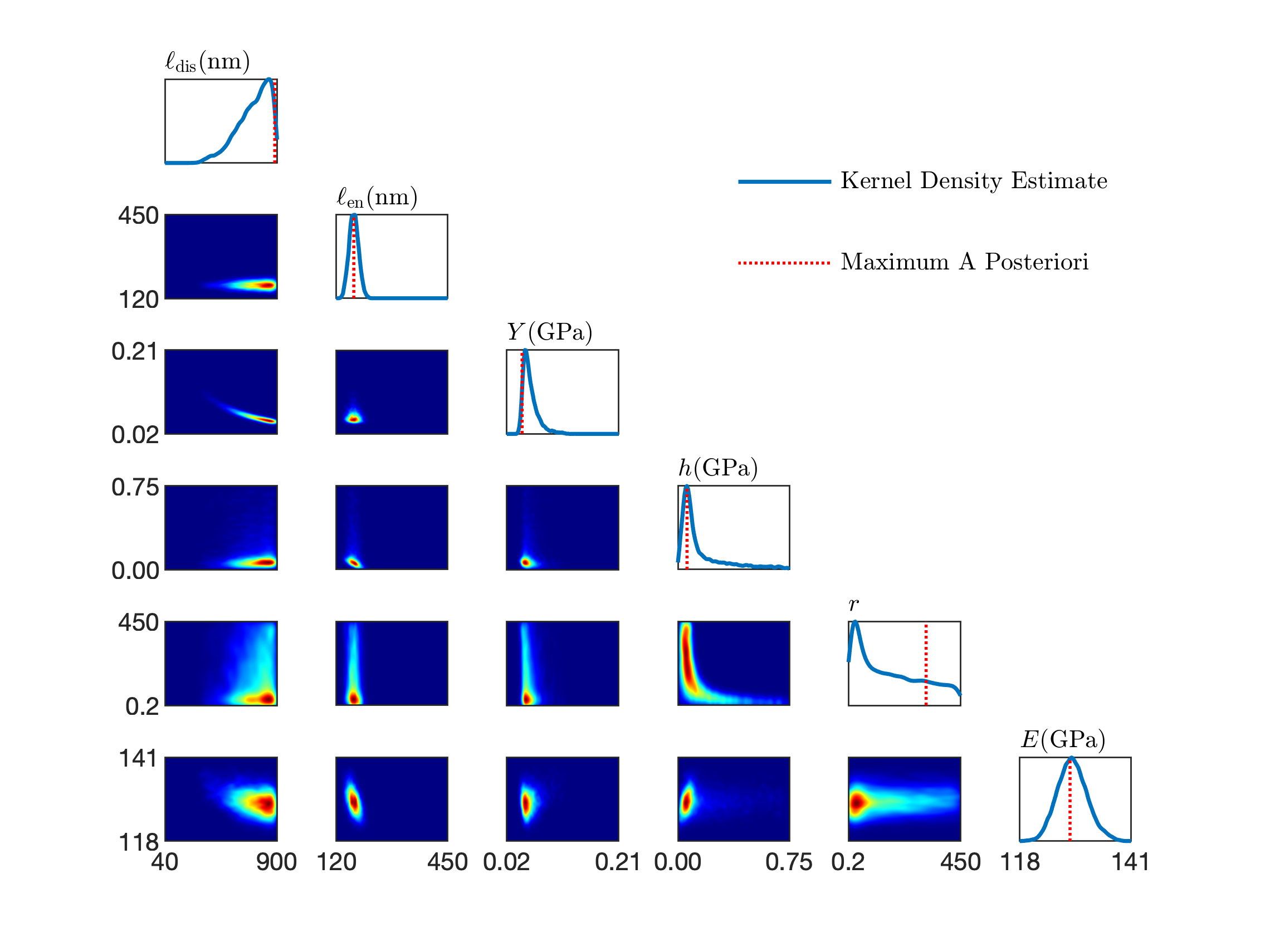}
\vspace{-0.25in}
\caption{The SGP’s parameter posteriors obtained from Case I of the Bayesian model calibration (see Table \ref{tabel:map}).
The training data sets consist of the MDDP simulations of 1000 nm, 700 nm, 500 nm, 300 nm micro-pillars.
The 1D plots represent the kernel density estimations (KDEs) of the marginal posterior distributions, and the dashed line is the MAP estimates of the parameters. The joint bivariate posteriors distributions are the 2D KDE plots.
The range of these plots corresponds to the parameter's uniform priors according to Table \ref{table:param}.
}
\label{fig:kdes}
\vspace{-0.1in}
\end{figure}

{
The kernel density estimations (KDEs) of the SGP's parameter posteriors are shown in Figure \ref{fig:kdes}. 
Both marginal distributions, corresponding to a single parameter, and joint bivariate distributions, corresponding to two parameters, are shown in this figure. The plots' range agrees to the uniform priors of the parameters presented in Table \ref{table:param}.
The MAP point estimates, $\boldsymbol{\theta}^{\text{MAP}}$, according to (\ref{eq:map}), and 
the normalized variance of the posteriors, $\mathcal{I} ({\theta}_k)$, given in (\ref{eq:info}), are estimated from the MCMC solutions and presented in Table \ref{tabel:map}. 
Figure \ref{fig:kdes}, 
indicates that most of the parameters are learned from the training data, with an average normalized variance of 0.35.
The parameters $\ell_{\text{en}}$ and $Y$ are informed better than the others,
judging by the sharp posterior PDFs and small values of $\mathcal{I}(\ell_{\text{en}})$ and $\mathcal{I}(Y)$.
However, the posteriors of the isotropic hardening parameters $h$ and $r$ (least sensitive parameters) show that we have lower confidence about these two parameters compared to the others after inference. 
Comparing the posteriors of $h$ and $r$ with $\ell_{\text{en}}$ indicates that the hardening rate observed in the synthetic data is mainly attributed to gradient-dependent kinematic hardening, while the isotropic hardening parameters are picked around small values. 
Additionally, the joint bivariate posterior distributions of $Y$ and $\ell_{\text{dis}}$ indicate a strong correlation among these parameters due to the constitutive relations of the dissipative stresses in (\ref{eq:dissipative}). The Bayesian calibration leads to high confidence in $Y$, while $\ell_{\text{dis}}$ (one of the most sensitive parameters) is less informed by the data leading to a 53.57\% difference in the values of normalized variance between the two parameters. Such results stem from the unrecognizable elastic to plastic transition regimes in the stress-strain data of the smaller micro-pillars (500 nm and 300 nm), leading to limited identification of the gradient-strengthening parameter, $\ell_{\text{dis}}$.
}
%

\begin{figure}[!ht]
\centering
\includegraphics[trim = 0mm 0mm 0mm 0mm, clip, width=.45\textwidth]{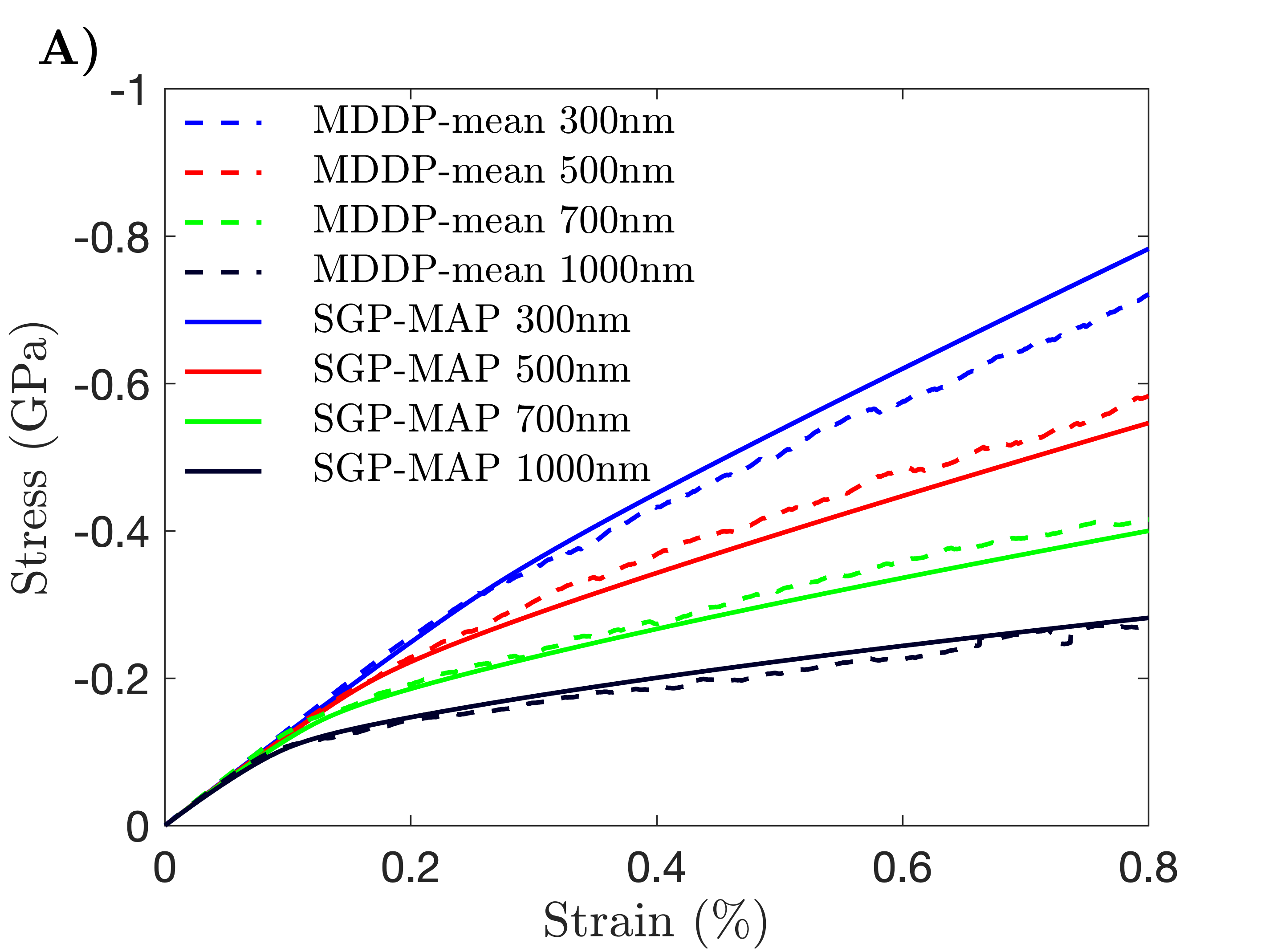}
~
\includegraphics[trim = 0mm 0mm 0mm 0mm, clip, width=.45\textwidth]{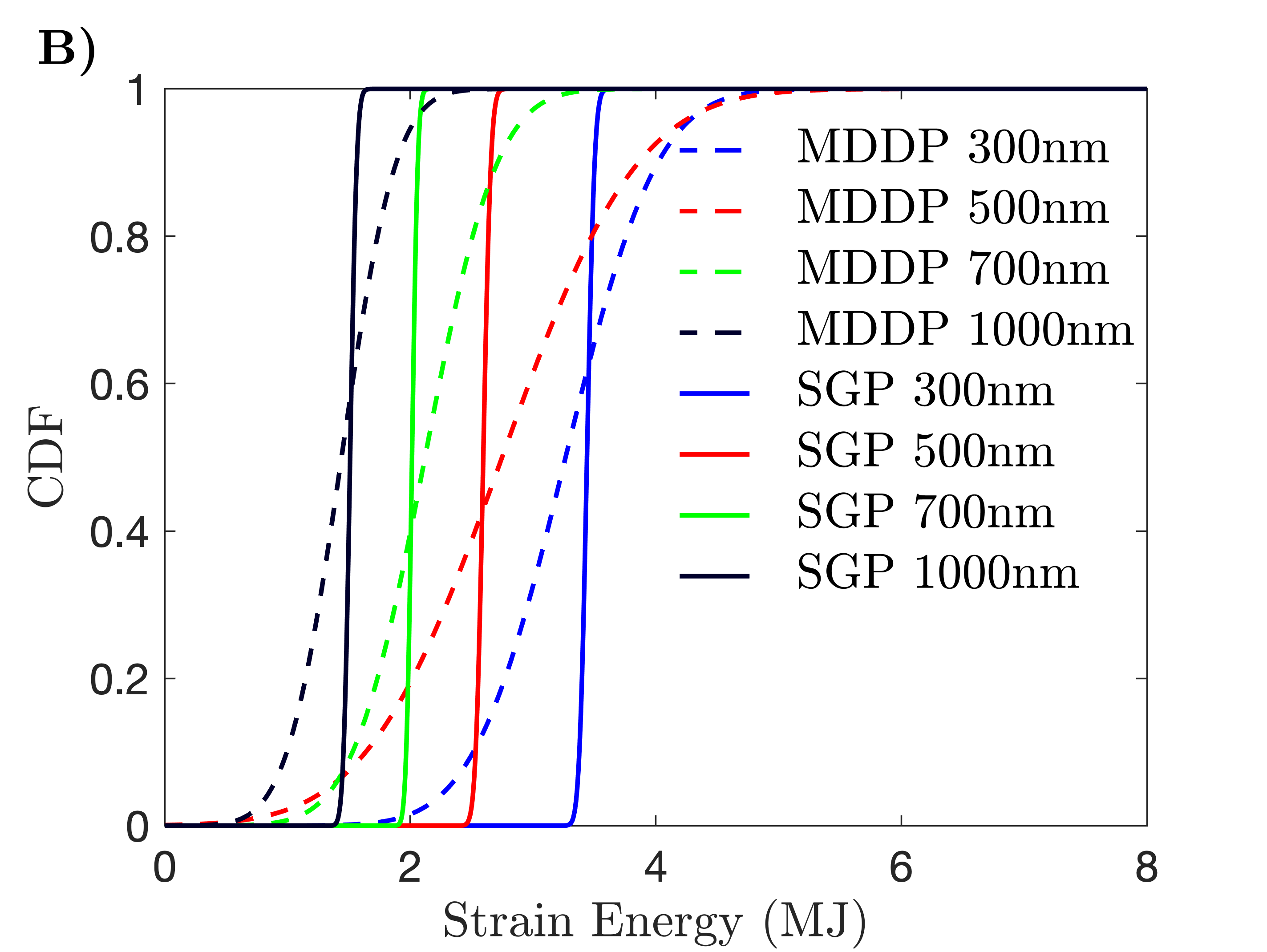}
~
\includegraphics[trim = 0mm 0mm 0mm 0mm, clip, width=.45\textwidth]{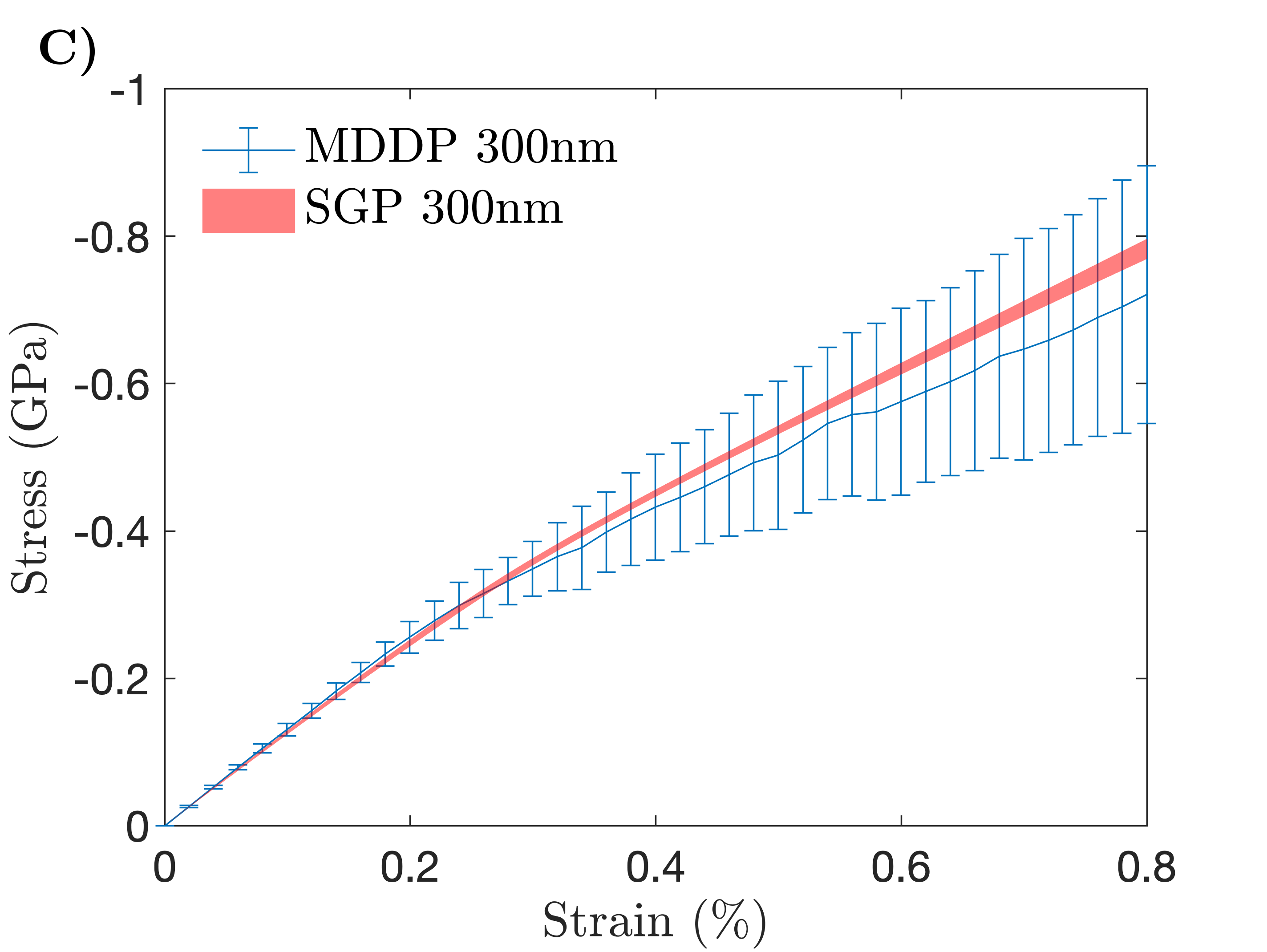}
~
\includegraphics[trim = 0mm 0mm 0mm 0mm, clip, width=.45\textwidth]{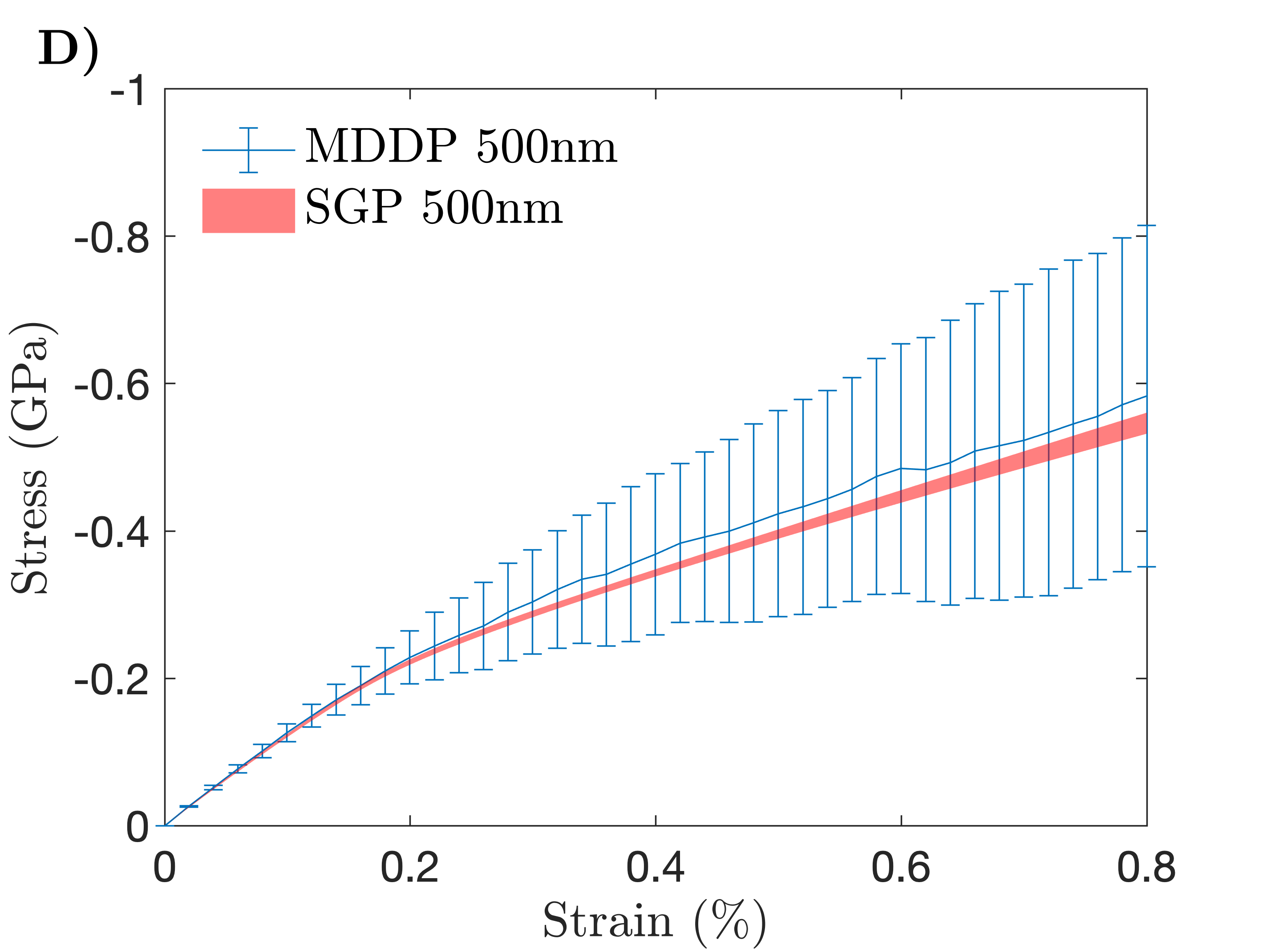}
~
\includegraphics[trim = 0mm 0mm 0mm 0mm, clip, width=.45\textwidth]{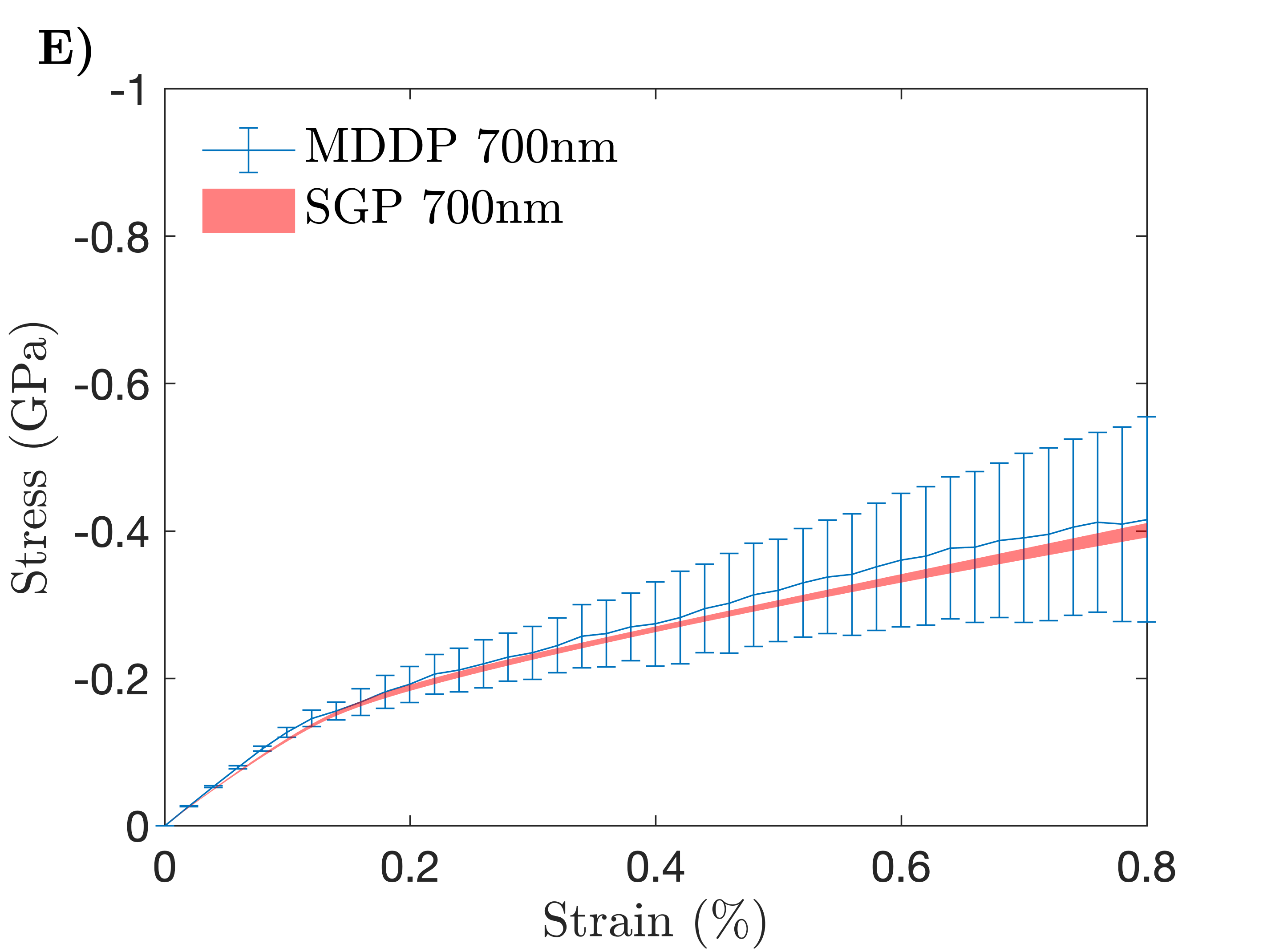}
~
\includegraphics[trim = 0mm 0mm 0mm 0mm, clip, width=.45\textwidth]{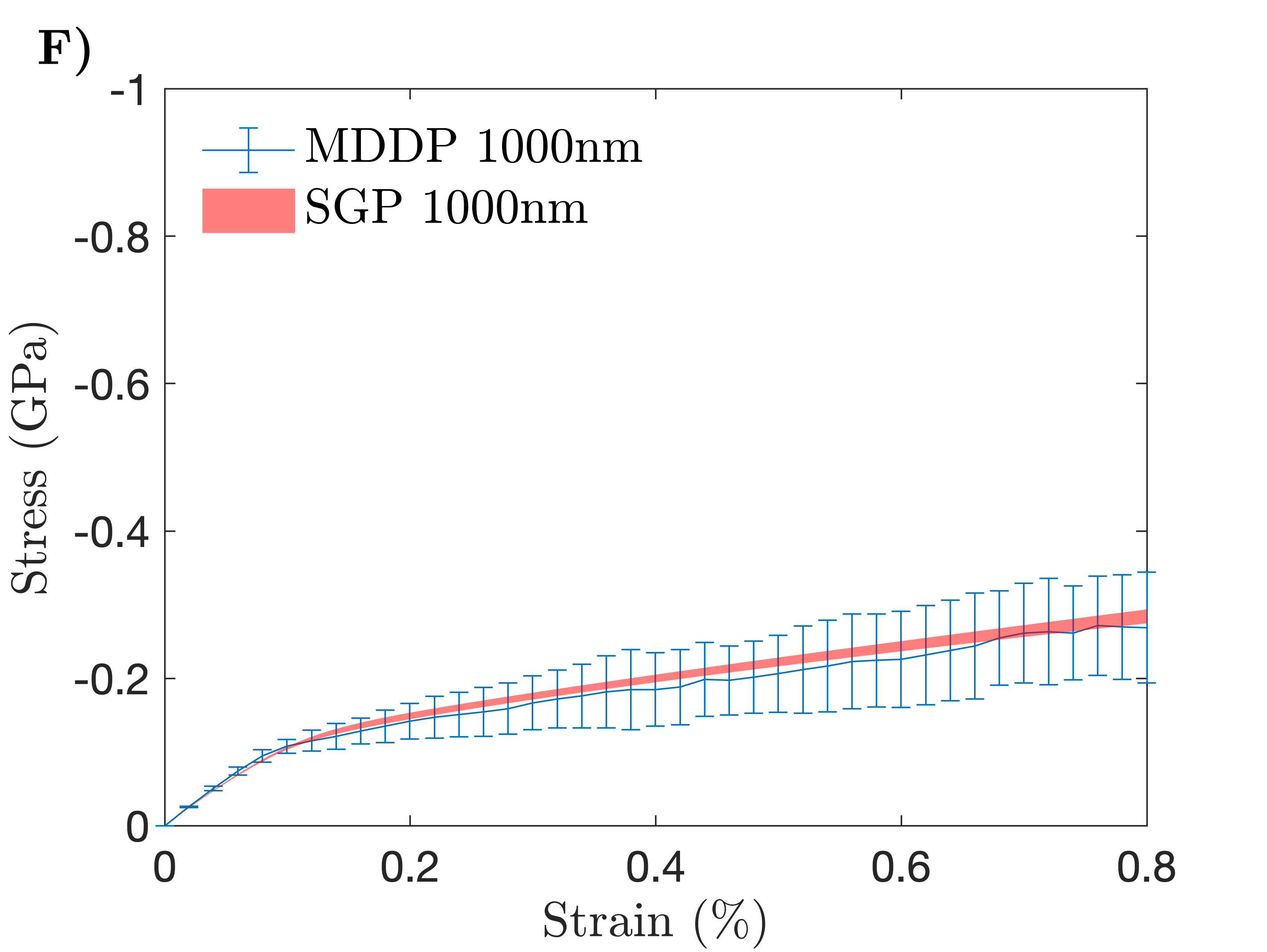}
\vspace{-0.05in}
\caption{
Comparison of the SGP model and MDDP simulations of micro-pillars for the training sets of Case I .
(A) comparing the stress-strain of the data means and the model prediction using the MAP points of the calibrated parameters, $\boldsymbol{\theta}^{\text{MAP}}$.
(B) comparing the cumulative distribution functions (CDF) of the strain energies (QoI), obtained from the MDDP data and the calibrated SGP model.
(C to F) comparing stress-strain of the MDDP data and SGP model predictions for the  300nm, 500nm, 700nm, and 1000nm micro-pillars. 
The means and standard deviations for the synthetic data are shown in blue, and the red area is the posterior prediction of the SGP model. 
For better presentation, the error bars are shown every 50 data points. 
}
\label{fig:calibration}
\vspace{-0.1in}
\end{figure}

Figure \ref{fig:calibration} compares the stress-strain and the strain energy (QoI) of the micro-pillars, obtained from the stochastic MDDP and SGP simulations, in the calibration scenarios (training sets).
This figure presents the MDDP and SGP calculations of the cumulative distribution functions (CDF) of the strain energies (panel B) and the stress-strain responses of the micro-pillars (panels C to F). The error in QoI, measured by (\ref{eq:error}), is presented in Table \ref{tabel:scenarios}.
These results indicate that the calibrated SGP model captures the size effect observed in the MDDP simulations in the 300 nm, 500 nm, 700 nm, and 1000nm micro-pillars with a 5.8\% of average error in strain energy.
Figure \ref{fig:calibration}(C to F), shows that the uncertainty in SGP computations of the stress-strain responses and the strain energy is much smaller than the data noise level in the MDDP simulations. 
Such high confidence in SGP model prediction, by part, is due to a large number of synthetic data points (approximately 2000 for each micro-pillar size) furnished by the MDDP simulations. 
These observations indicate that, despite the significant uncertainties in MDDP simulations, the SGP is informed well by the stress-strain training data of micro-pillars in Case I.
\begin{table}[!ht]
\centering
\caption{
{Maximum A Posteriori (MAP) estimate, $\boldsymbol{\theta}^{\text{MAP}}$, of the calibrated parameters according to (\ref{eq:map})
and the degree with which each parameter is learned from data, $\mathcal{I}(\boldsymbol{\theta})$, according to (\ref{eq:info}).
For each case, different training and testing data sets are considered (see Table \ref{tabel:scenarios}).}
}
\begin{tabular}{c|c|c|cc}
\hline
Calibration           & \multicolumn{2}{c|}{\textbf{Case I}}                                                                                          & \multicolumn{2}{c}{\textbf{Case II}}                   \\ \cline{2-5} 
Parameters          & $\boldsymbol{\theta}^{\text{MAP}}$     & $ \mathcal{I}(\boldsymbol{\theta}) $   & \multicolumn{1}{c|}{$\boldsymbol{\theta}^{\text{MAP}}$}         & $\mathcal{I}(\boldsymbol{\theta})$                  \\ \hline
$\ell_{\text{dis}}$  &	882.08 nm	&    0.254     &   \multicolumn{1}{c|}{502.23 nm}   &    0.452                  \\
$\ell_{\text{en}}$   & 	172.22 nm	&    0.021     &   \multicolumn{1}{c|}{108.28 nm} &    0.018                  \\
$Y$                       & 	0.046 GPa	&    0.055     &   \multicolumn{1}{c|}{0.078 GPa}   &    0.417                  \\
$h$                       & 	0.061 GPa	&    0.545     &   \multicolumn{1}{c|}{0.278 GPa}   &    0.414                  \\
$r$                        & 	311.68		&    1.040     &   \multicolumn{1}{c|}{257.43} &    0.662                  \\
$E$                       & 	128.45 GPa	&    0.213     &   \multicolumn{1}{c|}{130.20 GPa} &    0.202                 \\ \hline
\end{tabular}
\label{tabel:map}
\end{table}
%

\begin{figure}[!ht]
\centering
\includegraphics[trim = 0mm 0mm 0mm 0mm, clip, width=.48\textwidth]{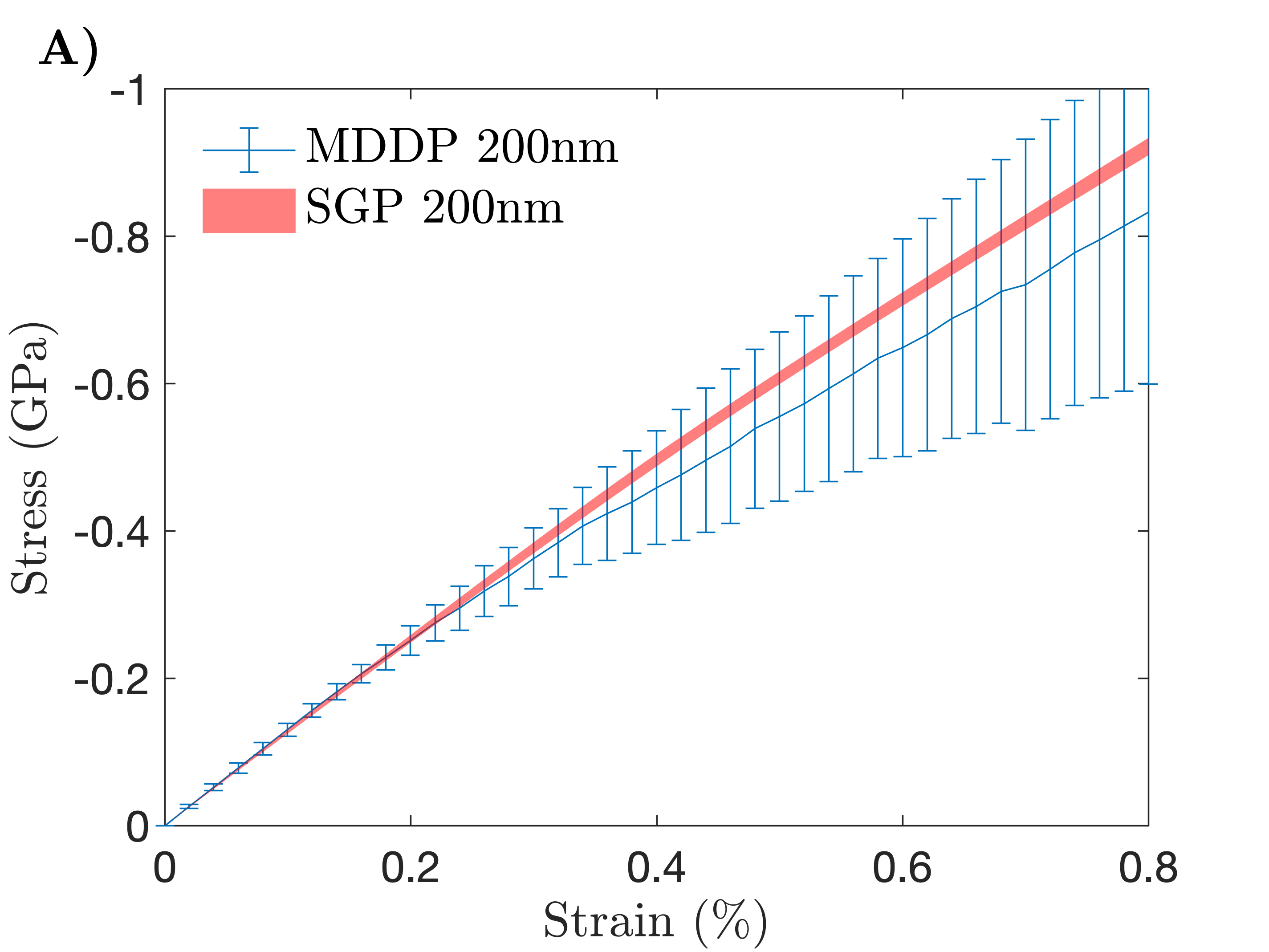}
~
\includegraphics[trim = 0mm 0mm 0mm 0mm, clip, width=.48\textwidth]{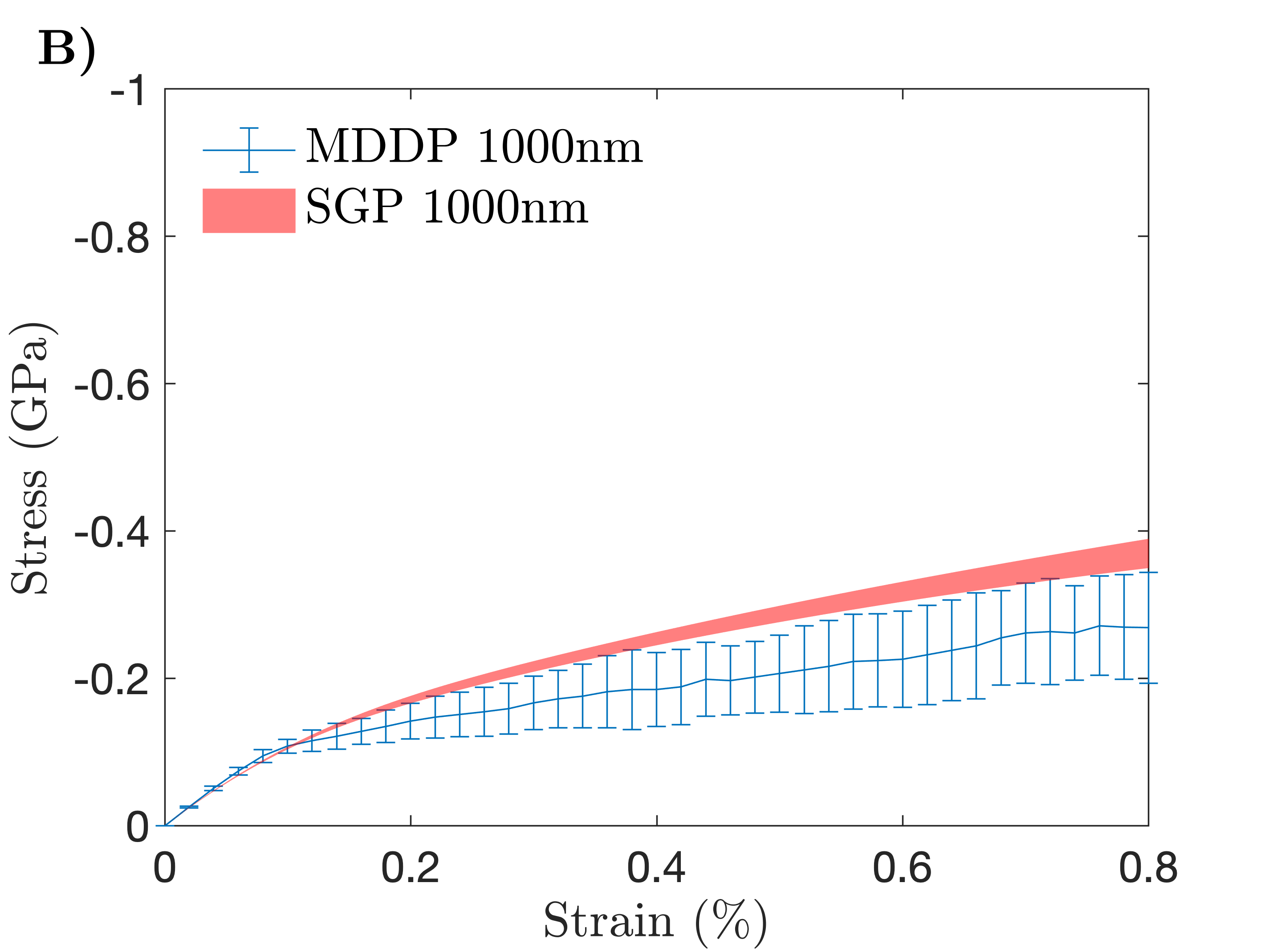}
\vspace{-0.15in}
\caption{
{
Comparison of the SGP model and MDDP simulations of micro-pillars for the testing sets of the Case I and Case II.
The stress-strain of the MDDP data and SGP model predictions of
(A) Case I (200 nm micro-pillar),
(B) Case II (1000 nm micro-pillar).
}
}
\label{fig:prediction}
\vspace{-0.1in}
\end{figure}

{
Next, we explore the calibrated SGP model's ability to predict the size effect behavior observed in MDDP simulations in the testing set (200nm micro-pillar), i.e., the stress-strain data excluded from the Bayesian calibration process.
Figure \ref{fig:prediction}(A) show the predicted stress-strain by the SGP model compared to the MDDP simulations of the 200nm micro-pillar. The error in predicting the target QoI by the SGP model in the testing set is below 10\% (see Table \ref{tabel:scenarios}). 
Additionally, the uncertainty in the SGP model prediction (average variance of stress = 0.0585 GPa) is remarkably lower than the noise level in the MDDP simulations of the 200 nm micro-pillar (average variance of stress = 0.7195 GPa), leading to a variance reduction of 91.87\% after Bayesian inference.
Overall, the Bayesian calibration of the SGP from the MDDP simulations in Case I indicates that the model parameters, except $r$, are learned from the synthetic data with a high confidence level. 
As a result, the stochastic SGP model captures the size effect in training sets and reliably predicts the unseen MDDP simulations in testing sets.
}

\subsubsection{Case II calibration and prediction}

{
According to Table \ref{tabel:scenarios}, the training data $\mathbf{D}$ in Case II consists of the MDDP stress-strain results obtained from the micro-pillars with the sizes of 700 nm, 500 nm, 300 nm, and 200 nm, and we aim to predict the responses of the 1000 nm micro-pillar.
The Bayesian calibration results indicate that, similar to Case I, the model parameters are informed well from the data in Case II. Additionally, the average error in QoI (Table \ref{tabel:scenarios}) in the training set of Case II shows a 28.85\% decrease compared to Case I, indicating that the SGP model in Case II better captures the size effect phenomena of MDDP simulations in the training sets.
However, the predicted stress-strain by the SGP model compared to the MDDP simulations of the 1000nm micro-pillar shown in Figure \ref {fig:prediction}(B) indicates that the calibrated model in Case II cannot accurately predict the unseen testing data set.
That is, the error in predicting the QoI of 1000 nm micro-pillar (prediction scenario) in Case II is above 30\%, indicating that the calibrated SGP model is invalid for computational prediction of the size effect plasticity.
Such outcomes are attributed to the stress-strain responses of MDDP simulations in different micro-pillar sizes. Due to the scarceness of dislocations in smaller micro-pillars with low dislocation density, the macroscopic yield points (elastic-plastic transition stress) are not apparent in the MDDP stress-strain results. Consequently, the synthetic data from larger micro-pillars are more informative for training the SGP model parameters, and excluding the MDDP simulations of the 1000 nm, micro-pillar from the calibration process reduces the predictive capability of the SGP model.
From these studies, one can conclude that adequate representation of the training data by the model does not necessarily result in a reliable model prediction. 
The computational prediction that goes beyond the available data requires an in-depth understanding of the underlying physics of the systems, along with taking advantage of physics-based computational models.
}

\section{Discussion and Conclusions}


{
This paper develops a sequential multiscale model from discrete MDDP simulations to a continuum SGP model to predict the size effect in plastic deformations of metallic micro-pillars under compression.
Comprehensive uncertainty analyses are conducted, including forward uncertainty propagation in SGP and Bayesian inference, to assess the reliability of multiscale model prediction.
The MDDP simulations of micro-pillar compression for different sizes and initial dislocation source heterogeneity indicate significant uncertainty in the stress-strain results. The MDDP studies also indicate that the primary deformation mechanism in the micro-pillar is the spiral motion of dislocations, i.e., half Frank-Read source, that tends to enclose the deformation in a small area leading to the significant sensitivity of the strength to the initial distribution of dislocations.
The variance-based global sensitivity analyses of the SGP show that the impact of the parameter uncertainty on the micro-pillars' strain energy (QoI) strongly depends on the size. The total sensitivity index of the elastic modulus decreases, while the effects of yield strength and isotropic hardening increase with the size of micro-pillars. These results stem from the size effect in stress-strain responses, in which the effect of classical plasticity parameters dominates the sensitivity of elastic properties as the micro-pillar sizes increase.
The sensitivity analysis also shows that the strain energy in 300 nm and 500 nm micro-pillars is highly sensitive to the variation in both energetic length scale (controlling gradient-dependent kinematic hardening) and dissipative length scale (regulating gradient-dependent flow stress). In comparison, the impact of the length parameters diminishes for larger (700 nm, 1000 nm) and smaller (200 nm) micro-pillars. Such responses are due to a reduction in the plastic strain gradient effect in larger domains and restriction of the boundary layer development for very small micro-pillars.
These results confirm that the length scales have a substantial effect if domain dimensions are of comparable magnitude; otherwise, their effects are weak.
Furthermore, the sensitivity analysis allows identifying the parameters that must be accurately calibrated and thus guides future experiments and DDD simulations to improve the SGP training and predictive capability.
For example, the stress-strain results, including loading and unloading, provide more information on distinguishing the kinematic and isotropic hardening responses of micro-scale materials. Consequently, the SGP parameters can be learned adequately from fewer synthetic or measurement data sets involving loading and unloading stages.
We note that the one-dimensional SGP modeling of the micro-pillar compression is very similar to the responses of thin films bonded to rigid substrates and subjected to pure shear loading. Thus, the conclusions on the parameter sensitivity of the SGP model can be extended to thin-film problems that are widely studied in the literature, e.g., \cite{faghihi2012thermal, fredriksson2005size}.}

{
The discrete to continuum multiscale is developed by calibrating the SGP model parameters using the synthetic data generated by MDDP simulations of micro-pillars.
A Bayesian calibration is employed to make prediction of the size effect in plastic deformation and quantify the uncertainty due to data noise and modeling errors. The data uncertainty originates from microstructural randomness in MDDP simulation due to different density and spatial distributions of dislocations and is the primary source of uncertainty in predictive modeling. The modeling error consists of simplifying assumptions in SGP constitutive relations to model complex dislocation interactions and the use of a one-dimensional model to simulate three-dimensional micro-pillar behavior.  
The statistical inference results indicate that the MDDP synthetic data in the micro-pillar with sizes 300 nm, 500 nm, 700 nm, and 1000nm (training sets) adequately inform the SGP model parameters. The calibrated SGP reliably predicts the MDDP simulation of 200 nm micro-pillar (testing set), with an error below 10\%. Moreover, despite the significant uncertainties in MDDP simulations, the Bayesian calibration leads to remarkably high-confidence in SGP's computational predictions.
Interestingly, excluding the MDDP simulations of the 1000 nm micro-pillar from the calibration process leads to the SGP better fit the MDDP training data while limiting the predictive capacity of the SGP model. 
Due to the scarcity of dislocations, the macroscopic yield points are not recognizable in stress-strain results of smaller micro-pillars; thus, the SGP model parameters are learned more from the MDDP simulations of the 1000 nm and 700 nm micro-pillars.
}


{
Several studies attempt to calibrate nonlocal continuum plasticity models from either DDD simulations, e.g., \cite{shu2001, GiessenNeedleman1997, chang2012}, or micro-scale experimental data, e.g., \cite{meng2014micro, faghihi2014jemt, fleck1994strain} to predict size effect plasticity in microscale metallic materials.
These investigations have enabled understanding the physical origin of the material intrinsic length scale underlying the size effect and refining SGP models to capture material responses accurately. 
However, in all these efforts, the inherent stochasticity in measurement or synthetic data is treated by ensemble averaging. To the best of our knowledge, the current paper is the first attempt to account for the randomness in DDD's microstructural evolutions, quantify the model and data uncertainties, and assess the credibility of size effect prediction of the SGP models. 
The comprehensive uncertainty analyses of the SGP model, introduced in this contribution, enable taking advantage of these models in computational prediction of novel material systems on practical time and length scales.
Furthermore, the presented predictive modeling framework is readily transferable to other discrete-continuum models of other material systems and allows for uncertainty characterization in multiscale models' predictions.
}


{
Despite the comprehensive uncertainty analyses of the DDD-SGP multiscale model in this study, several areas can be addressed in future studies. 
A critical aspect of predictive computational modeling is assessing the validity of the model. According to \cite{odenbabuska2017}, the Bayesian model validation requires additional (possibly more complicated) scenarios that provide validation data for re-calibrating the model. The model prediction accuracy is then tested against a validation tolerance to determine if the model is valid.
The model validation process is critical to advance the DDD-SGP multiscale models' ability to predict macro-scale material and structural systems.
Moreover, the material parameters of the MDDP simulations in the current study, such as elastic properties and dislocation mobility, are assumed to be the same in all the micro-pillars. One can consider these parameters as random variables and account for additional source uncertainty in discrete simulations to represent more realistic material responses. 
Additionally, in this paper, the SGP model is calibrated using the stress-strain responses of the MDDP simulations. A more robust predictive multiscale model can be developed using stress-strain and the local distribution of plastic strain provided by discrete dislocation simulations to inform the SGP model. Bayesian methods provide a natural framework for the simultaneous use of multiple sources of data for parameter inference.
To this end, one can investigate whether the SGP is valid for predicting full features of discrete dislocation dynamics, including boundary layer development that gives rise to the size effect phenomena.
Finally, we used a one-dimensional SGP model of the micro-pillar compression for the UQ analyses since performing hundreds of thousands of model evaluations of the three-dimensional finite element model is computationally infeasible. 
Such simplification may leave out crucial three-dimensional effects at the free boundaries depending on the micro-pillars' dimensions and the values of SGP model parameters. 
Additional numerical experiments using a two-dimensional SGP model indicate that for the range of the model parameters (Table \ref{table:param}) and the aspect ratio of the micro-pillars used in the current study, the boundary effect on stress-strain results are negligible. The mean error in the stress computed from one- and two-dimensional simulations is $< 2.5\%$, while a higher discrepancy among the two results is observed in the local distribution of plastic strain.
While our Bayesian inference accounts for such modeling errors, future investigation is required to assess the predictive capability of the three-dimensional SGP models of micropillars, specifically when the plastic strain profile is taken into account as an additional observable for parameter inference.
}

%
%

{
In conclusion, the outcome of this study indicates that the developed MDDP-SGP multiscale model can accurately simulate the size-dependent plastic deformation in microscale materials such as micro-pillars. Remarkably, rigorous characterization of microstructural randomness and modeling error shows that the SGP model can reliably predict the size effect plasticity responses of the micro-pillar outside of the training data range, despite the considerable variance in the MDDP simulations.
This study shows that the essential requirements in developing predictive discrete-continuum multiscale models are 
(i) designing scenarios of discrete simulations to furnish informative training data sets for calibrating the continuum model;
(ii) calibrating the continuum model with discrete simulation data while coping with uncertainties in model parameters and the stochasticity of the discrete model, which translates into uncertainties in model predictions.
}

\section*{Acknowledgments}
We benefited from the discussions with Dr. Kathryn Maupin, of the Sandia National Laboratories, on the software applications and the Bayesian inference solution.
{We are grateful to the referees for their constructive inputs.}

\bibliographystyle{apalike}
\bibliography{refs}
\index{Bibliography@\emph{Bibliography}}%

\end{document}